\documentclass[aps,prl,twocolumn,superscriptaddress]{revtex4-1}

\usepackage{amsmath}
\usepackage{amssymb}
\usepackage{braket}
\usepackage{graphicx}
\usepackage{textcomp}
\usepackage{units}

\begin{document}

\title{Coherent control and wave mixing in an ensemble \\ of silicon vacancy centers in diamond}

\author{Christian Weinzetl}
\affiliation{Clarendon Laboratory, University of Oxford, Parks Road, Oxford OX1 3PU, United Kingdom}

\author{Johannes G\"{o}rlitz}
\affiliation{Naturwissenschaftlich-Technische Fakult\"{a}t, Fachbereich Physik, Universit\"{a}t des Saarlandes, Campus E 2.6, 66123 Saarbr\"{u}cken, Germany}

\author{Jonas Nils Becker}
\affiliation{Clarendon Laboratory, University of Oxford, Parks Road, Oxford OX1 3PU, United Kingdom}

\author{Ian A. Walmsley}
\affiliation{Clarendon Laboratory, University of Oxford, Parks Road, Oxford OX1 3PU, United Kingdom}

\author{Eilon Poem}
\affiliation{Department of Physics of Complex Systems, 
Weizmann Institute of Science, Rehovot 76100, Israel}

\author{Joshua Nunn}
\affiliation{Centre for Photonics and Photonic Materials, Department of Physics, University of Bath, Claverton Down, Bath BA2 7AY, United Kingdom}

\author{Christoph Becher}
\affiliation{Naturwissenschaftlich-Technische Fakult\"{a}t, Fachbereich Physik, Universit\"{a}t des Saarlandes, Campus E 2.6, 66123 Saarbr\"{u}cken, Germany}

\date{\today}

\begin{abstract}
Strong light-matter interactions are critical for quantum technologies based on light, such as memories or nonlinear interactions. Solid state materials will be particularly important for such applications, because of the relative ease of fabrication of components. Silicon vacancy centers (SiV$^{-}$) in diamond feature especially narrow inhomogeneous spectral lines, which are rare in solid materials. Here, we demonstrate resonant coherent manipulation, stimulated Raman adiabatic passage, and strong light-matter interaction via four-wave mixing of a weak signal field in an ensemble of SiV$^{-}$ centers.
\end{abstract}

\maketitle

During the past decade, color centers in diamond have emerged as important systems for quantum information processing (QIP) \cite{Monroe2002, Kok2007}, as well as for quantum sensing and metrology applications \cite{Degen2017,Bernardi2017}. While the nitrogen vacancy center (NV$^{-}$) has dominated recent research \cite{Doherty2013} its prominent phonon sidebands (Debye-Waller factor, \mbox{$\text{DWF} \approx 0.04$}) \cite{Aharonovich2011} severely limit rates in probabilistic entanglement schemes and hence its applicability in QIP \cite{Hensen2015}. Recently, the negatively-charged SiV$^{-}$ emerged as a promising alternative system, offering remarkable optical properties such as a dominant zero-phonon line fluorescence at 737nm \mbox{($\text{DWF} \approx 0.8$)} \cite{Neu2011} and narrow inhomogeneous distributions of the SiV$^{-}$ resonance of single centers in the range of only a few hundred MHz for samples of low emitter density \citep{Sipahigil2014}. In combination with efficient coupling to diamond nanostructures \cite{Riedrich-m2012, Sipahigil2016} and microwave- \cite{Pingault2017} as well as all-optical coherent control \citep{Becker2016, PhysRevLett.120.053603}, this potentially allows the realization of highly-efficient coherent spin-photon interfaces \cite{Togan2010} using single SiV$^{-}$ centers.
\begin{figure}[t]
	\includegraphics[width=\linewidth]{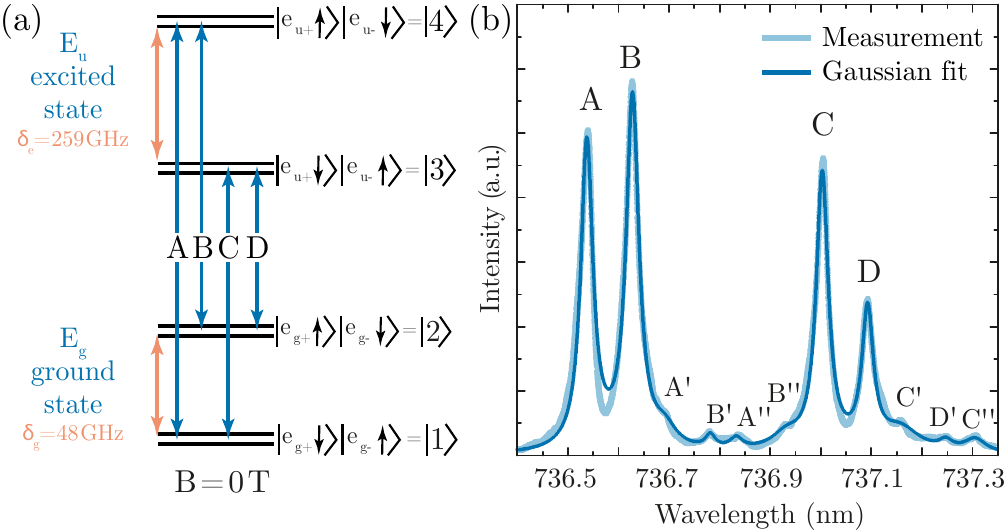}
	\caption{SiV$^{-}$ four-level structure \& optical zero-phonon line spectrum. (a) The four optical transitions A, B, C, D between the four orbital states $\ket{1...4}$ form a double-$\Lambda$ system. (b) Photoluminescence excitation spectrum at $\text{T}=5$\,K of the SiV$^{-}$ ensemble used throughout this work (light blue) \cite{Arend2016}. A fit of the spectrum to a set of Gaussian spectral lines indicates an inhomogeneous broadening of $\approx 10$\,GHz of transitions A, B, C, D (dark blue). Grown with a natural silicon isotope distribution, the sample displays twelve optical transitions (D'' transition not included in measurement).}
	\label{fig1}
\end{figure}

The electronic structure of the center is displayed in Fig.\,\ref{fig1}(a) and consists of two-fold spin-degenerate  orbital doublet ground and excited states in which the orbital components are split apart by strong spin-orbit (SO) interactions, leading to splittings of $\delta_{g}$=48\,GHz in the ground and $\delta_{e}$=259\,GHz in the excited state \cite{Hepp2014, gali2013}, with spin and orbital coherence times limited to \mbox{T$_2^* \approx 40$\,ns} at liquid helium temperatures by phonon-induced dephasing processes \cite{Pingault2014,Rogers20142,Jahnke2015}. Ground state spin coherence times can be increased to milliseconds at sub-kelvin temperatures \cite{PhysRevLett.120.053603,Sukachev2017}. At zero magnetic field and at cryogenic temperatures this level structure results in a characteristic four-line fine structure of the zero-phonon line (ZPL) \cite{Neu2013}, forming an optically accessible orbital double-$\Lambda$-type system, enabling a multitude of optical coherent control schemes \cite{Alzetta1976, Gaubatz1990, nunn2007}. The SiV$^{-}$ is a trigonal-antiprismatic complex featuring a silicon atom in a split-vacancy configuration in between two empty carbon lattice sites and six nearest-neighbour carbon atoms \cite{Goss1996, gali2013}. The resulting inversion symmetry renders the system insensitive to first-order Stark shifts \cite{Sipahigil2014}, enabling narrow spectral distributions of centers even in non-ideal crystal environments. Even in dense ensembles \cite{Sternschulte1994}, the  inversion symmetry  of SiV$^{-}$ centers offers an inhomogeneous broadening in the low GHz regime \cite{Arend2016}. Such narrow inhomogeneous linewidths are not achievable with other solid-state quantum systems such as NV centers \cite{Acosta2009, Poem2015}, quantum dots \cite{Suzuki2016}, or rare-earth-doped crystals \cite{Liu2005}. This allows for fundamental studies and applications in the fields of strong coherent light-matter interactions and single photon nonlinearities, which so far have been reserved for systems such as single atoms coupled to optical cavities \cite{Duan2004}, ultra-cold Rydberg-atoms \cite{Dudin2012}, or cold as well as hot atomic vapors \cite{Lukin2003, Reim2010}. In contrast, SiV$^{-}$ ensembles offer high photon interaction cross-sections even in small sample volumes and thus allow for the observation of similar effects in compact and scalable devices in the solid state, suitable for integration with waveguides and on-chip photonic structures.

In this Letter, we present a suite of coherent manipulations along with strong, collectively-enhanced light-matter coupling in a mesoscopic SiV$^{-}$ ensemble. The orbital coherence allows for fast manipulation, even at zero magnetic fields \cite{Poem2015}, where we capitalize on previous work on all-optical ultrafast coherent control of single SiV$^{-}$ centers \cite{Becker2016, Zhou2017, PhysRevLett.120.053603}. Specifically, we here demonstrate resonant Ramsey interference, optical Hahn echo, and stimulated Raman adiabatic passage (STIRAP) in SiV$^{-}$ ensembles. Furthermore, we observe strong light-matter interactions in the form of absorption and amplification of weak coherent fields by four-wave mixing (FWM).

In Fig.\,\ref{fig1}(b) a photoluminescence excitation (PLE) measurement of the SiV$^{-}$ ensemble used in this work is shown. The ensemble consists of a $\sim$300\,nm thick SiV$^{-}$-doped layer homoepitaxially grown on top of a pre-selected type Ib high-temperature-high pressure (HPHT) diamond substrate using a silicon source with a natural isotope distribution resulting in an ensemble displaying twelve optical transitions \cite{Sternschulte1994, Arend2016}. The main lines A-D correspond to SiV$^{-}$ centers containing $^{28}$Si, while energetically shifted transitions correspond to centers containing $^{29}$Si (A'-D') and $^{30}$Si (A''-D''), with intensity ratios matching the natural isotope abundances \cite{Sternschulte1994, Dietrich2014}. The Gaussian inhomogeneous linewidth of each main optical transition of this sample is $\approx 10$\,GHz, only about a factor of 100 larger than the natural linewidth of single SiV$^{-}$ centers and smaller than $\delta_g$, allowing for ultrafast manipulation of individual optical transitions even without the application of electric or magnetic fields.

To model coherent interactions in such an ensemble of emitters we developed an optical Bloch equation (OBE) model. The temporal evolution is governed by the master equation in Lindblad form,
\begin{equation}
\dot{\rho} = - \frac{\textsl{\textrm{i}}}{\hbar} \left[\mathcal{H}, \rho \right] + \sum^{4}_{i,j=1}\Gamma_{ij} (\rho_{ii}\ket{j}\bra{j}-\frac{1}{2} \{\ket{i}\bra{i},\rho \}) ,
\label{eq1}
\end{equation}
where $\mathcal{H}$ is the orbital four-level interaction Hamiltonian in rotating wave approximation, $\rho$ is the density matrix of a single SiV$^{-}$ center, and where we have introduced the spontaneous decay rates $\Gamma_{ij}$ ($i\neq j$) and pure dephasing rates $\Gamma_{ii}$, following \cite{Becker2016}. The coherent evolution of the ensemble was described by numerically solving Eq.\,(\ref{eq1}) for a number of individual SiV$^{-}$ centers, each with a different resonance frequency, i.e. different detuning relative to the applied electromagnetic fields, to account for the inhomogeneous broadening defined by a Gaussian distribution. The solutions for the individual emitters are then weighted by the amplitude of the Gaussian at the respective detuning and averaged.
\begin{figure}[t]
	\includegraphics[width=0.95\linewidth]{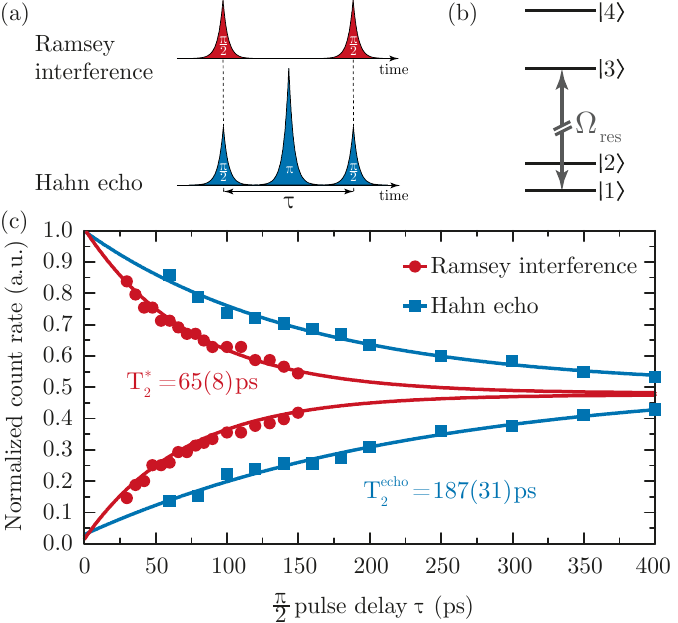}
	\caption{(a) Pulse sequence of 12\,ps long optical pulses for Ramsey interference (red, upper) and Hahn echo (blue, lower) measurements. (b) SiV$^{-}$ level system with a resonant optical field on transition C identical for both measurements. (c) Optical Ramsey interference (red circles, inner) and Hahn echo measurement (blue squares, outer). Measured envelopes (dots and squares) and simulations (solid lines) using a four-level OBE model assuming an ensemble of 10 emitters with a  10\,GHz broad Gaussian distribution are shown (T=5\,K, P$_\frac{\pi}{2}=0.5(2)\,\mu$W, P$_\pi=1.8(3)\,\mu$W). The pulse areas were experimentally optimized.}
	\label{fig2}
\end{figure}

In order to validate this theoretical model, resonant Ramsey interference measurements were performed using a sequence of two 12\,ps long optical pulses with variable delay $\tau$ resonantly applied to transition C of the ensemble, preparing a coherence between the lower ground state $\ket{1}$ and the lower excited state $\ket{3}$ of the ensemble. The pulse sequence and the addressed optical transitions are given in Fig.\,\ref{fig2}\,(a) and (b). The fluorescence intensity as a function of the delay $\tau$ is measured, which reflects the upper state population after the pulse sequence. In Fig.\,\ref{fig2}\,(c) individual fringes oscillating at optical frequencies were not resolved and only the upper and lower envelopes are shown for both measurements. With an inhomogeneous broadening of 10\,GHz inferred from Fig.\,\ref{fig1}(b) and the transition rates measured for a single emitter in \cite{Becker2016} as input parameters, the OBE ensemble model appropriately describes the resulting Ramsey fringe decay averaging over only a small number of ten emitters in the model and thus limiting its numerical complexity. An excitation-induced dephasing of about 450\,MHz, most-likely related to phonon broadening by the applied laser pulses, has to be added to match the measured coherence decay. Similar effects have already been observed for single SiV$^{-}$ centers using ultrafast laser pulses \cite{Becker2016}. The observed fringe decay indicates an inhomogeneous coherence time of T$^*_2$ = 65(8)\,ps, about a factor of 15 lower than what has been observed for single centers \cite{Becker2016}. This is caused by a relative dephasing of individual emitters in the ensemble with different Larmor frequencies due to their varying detuning relative to the applied optical field.
\begin{figure}[t]
	\includegraphics[width=0.95\linewidth]{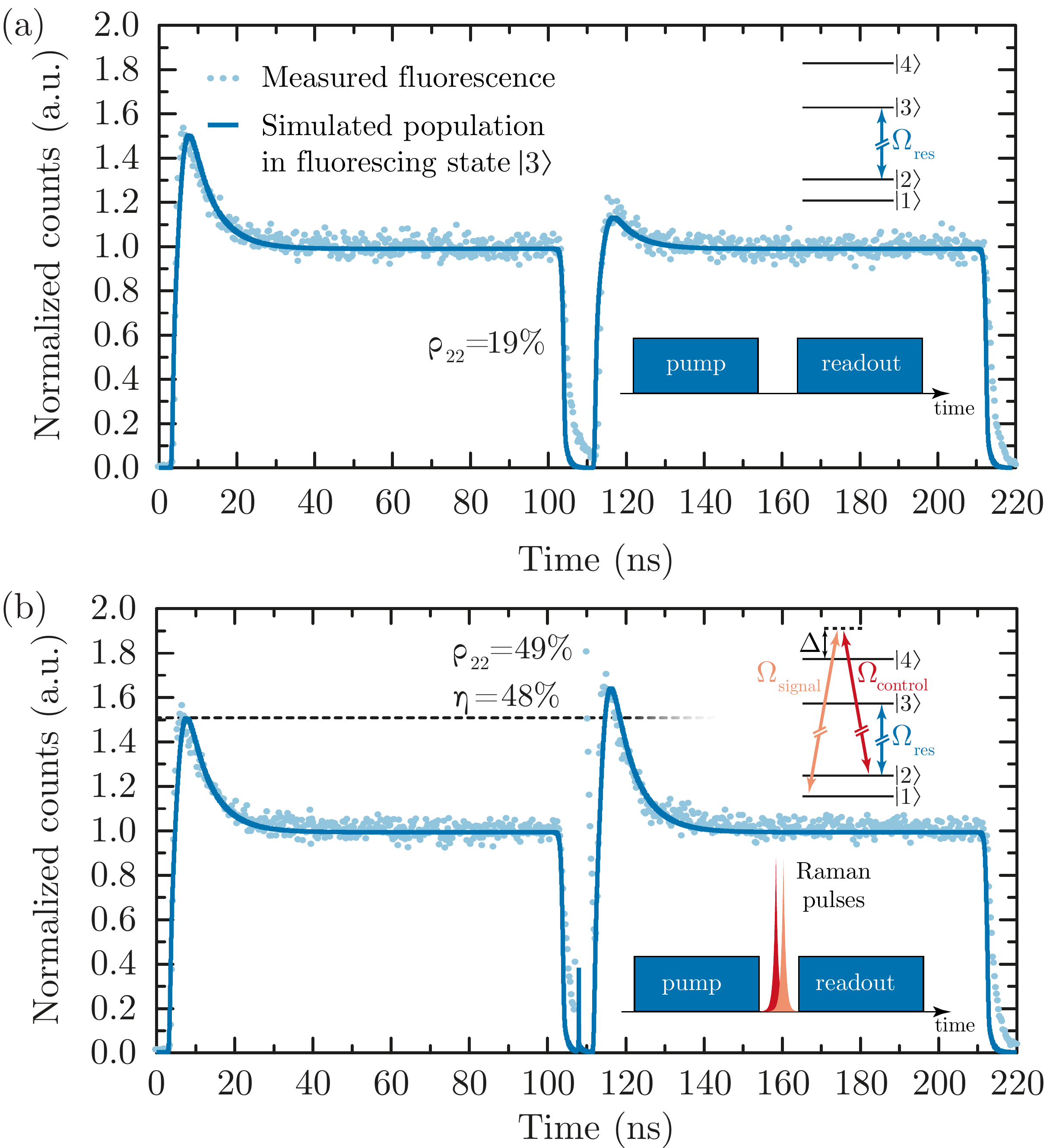}
	\caption{Stimulated Raman adiabatic passage. The fluorescence into the phononic sidebands is measured both (a) without and (b) with the Raman pulses between the pump and readout pulses (light blue dots). For simulations the OBE model is used (blue line). (a) Optical pumping and readout of the ground orbital state $\ket{2}$ using resonant illumination of transition D. The OBE model indicates a population in $\ket{2}$ of $\rho_{22} = 19 \%$ in between both resonant pulses. (b) Sequence from (a) including the Raman signal and control field applied in between the resonant pulses with zero temporal delay to each other. Both fields are blue-detuned by \mbox{$\Delta$ = 70\,GHz} from the upper excited state $\ket{4}$ while in two-photon resonance. A population of $\rho_{22} = 49 \%$ equivalent to a transfer efficiency of $\eta = 48 \% $ is reached directly after the Raman pulses have been applied. Inserts show optical field configurations and pulse sequences. The power of the resonant, signal, and control fields are P$_{\text{res}} = 30\,\mu$W, P$_{\text{signal}} = 33\,\mu$W, and P$_{\text{control}} = 27\,\mu$W respectively.}
	\label{fig3}
\end{figure}

Using an optical Hahn echo sequence, we explored the possibility of counteracting this dephasing which is caused by local variations of the static strain field in the sample. The measured Hahn echo envelopes result in an improved coherence time of T$_{2}^{\text{echo}}$=187(31)\,ps. An excitation-induced dephasing of about 900\,MHz needed to be added, in accordance to the doubled total pulse area in comparison to the Ramsey interference pulse sequence. Our model indicates that, in principle, values equivalent to the single-emitter coherence time can be reached with this technique. Albeit dephasing can be suppressed further by optimizing thermal anchoring, the observed increase in coherence time outlines the great potential of similar echo sequences in future experiments to mitigate the effects of the remaining inhomogeneous broadening.

This set of experiments demonstrate coherent manipulation of the ensemble involving its excited state, however, for many applications it is desirable to access the longer-lived and more narrowband ground state coherence. This coherence can be addressed optically using off-resonant two-photon Raman transitions for which the ground state coherence time of the ensemble can be estimated to \mbox{T$_{2\text{,g}}^* \approx 100$\,ps}, in conformity with coherent population transfer (CPT) measurements \cite{Arend2016, supp}. Here, we first demonstrate the feasibility of ensemble Raman transitions by inducing a coherent population transfer between the two ground states $\ket{1}$ and $\ket{2}$ by implementing STIRAP using ultrafast off-resonant signal and control pulses \cite{Gaubatz1990, Vitanov2017}. For this, initialization of the system in $\ket{1}$ and readout of the resulting population of $\ket{2}$ is realized by means of optical pumping, resonantly driving transition D with two subsequent 100\,ns long pulses as displayed in Fig.\,\ref{fig3}\,(a), reducing the population $\rho_{22}$ in $\ket{2}$ from its thermal equilibrium value of $\rho_{22}$=37\% at 5\,K to a minimum of $\rho_{22}$=19\%, 6\,ns after the end of the pump pulse, limited by the competition between optical pumping and relaxation between the ground states with a time constant of T$_1^\text{orbit}$ = 27(8)\,ns \cite{supp}. With the same delay from the pump pulse, in between the two resonant pulses, the two Raman pulses are then applied, leading to a coherent population transfer from $\ket{1}$ to $\ket{2}$ as indicated by the strong increase of the readout pulse rising edge fluorescence in Fig.\,\ref{fig3}\,(b). This peak corresponds to a population in $\ket{2}$ of $\rho_{22} = 49 \% $. All population values are inferred from the OBE model. The model further reveals that the population transfer in the limit of ultrashort Raman pulses and zero relative pulse delay is bidirectional between $\ket{1}$ and $\ket{2}$ with the same transfer efficiency $\eta$ such that in the case of perfect efficiency the populations in the ground states are swapped. Therefore, the final population $\rho^f_{22}$ after STIRAP is limited by the initial population $\rho^{i}_{22}$ in $\ket{2}$ given by imperfect initialization. Whereas the simulation predicts a transfer efficiency maximum close to 100\% \cite{supp}, in the experiment, a transfer efficiency of $\eta$ = $\nicefrac{(\rho^{f}_{22}-\rho^{i}_{22})}{(1 - 2\rho^{i}_{22})}$  = 48\% is reached. According to the model, this is mainly limited by the available power in the Raman pulses. The population transfer via two-photon transitions was confirmed by applying the control and signal fields separately. No population transfer was observed by such single pulses. Furthermore, population transfer was suppressed by tuning the fields out of two-photon resonance and increasing their temporal delay as expected for a two-photon process \cite{supp}.

\begin{figure}[t]
	\includegraphics[width=1\linewidth]{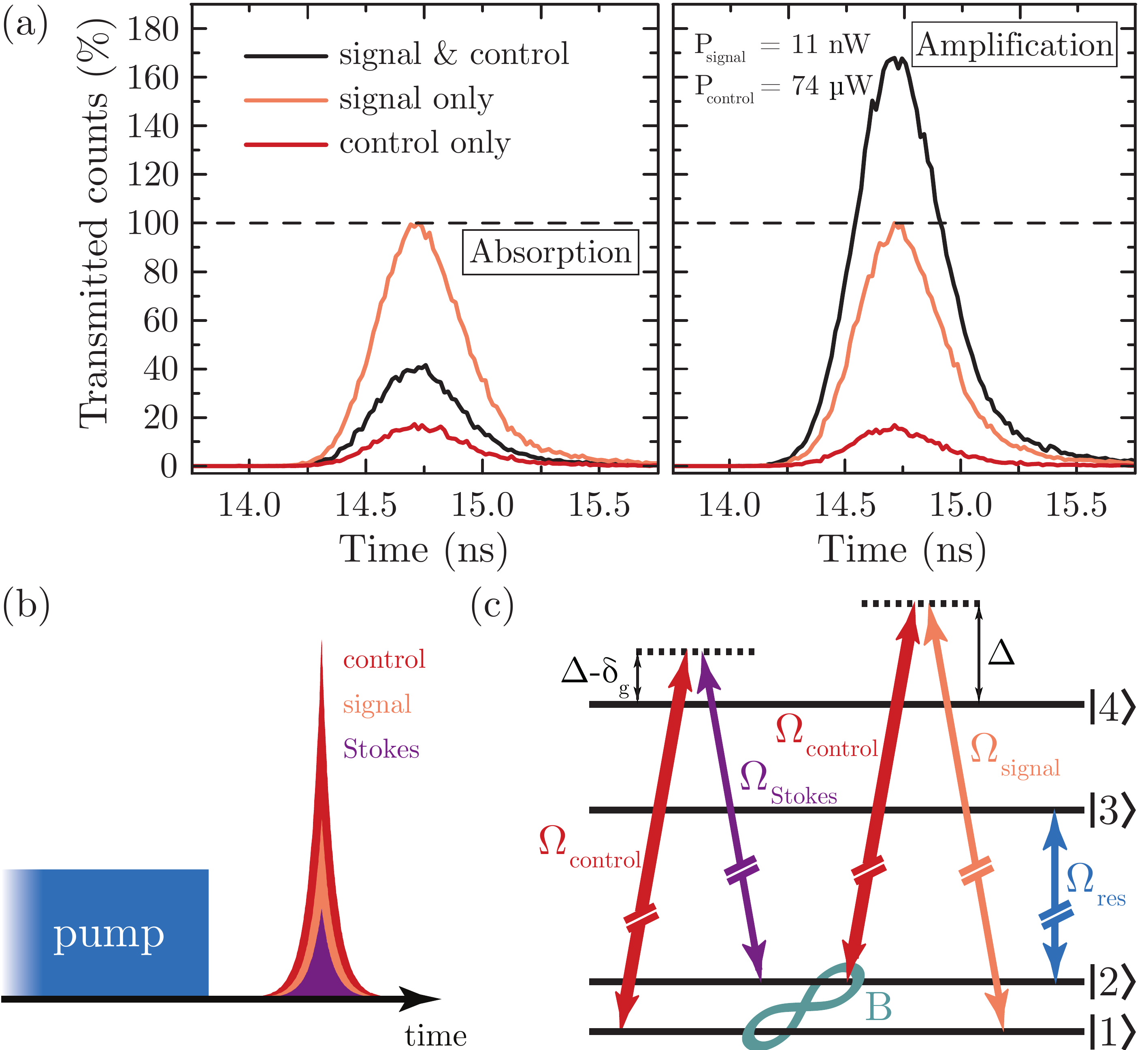}
	\caption{Signal absorption and amplification. (a) Time resolved measurement of optical fields sent through the ensemble. The filter setup is set to maximum transmission for the signal wavelength. The fields are normalized in amplitude to the counts measured when only the signal field is present. Subtracting the background control field, which could not be entirely suppressed by the filter setup, $\approx 80$\% absorption (left) and $\approx 60$\% amplification (right) of the signal field were measured. (b) Pulse sequence. After preparing the ensemble by resonant optical pumping, control, signal, and Stokes field are applied simultaneously. (c) The strong control field couples to both ground states creating two $\Lambda$-systems with different detuning from the excited states. Stokes and signal field which are also applied to the sample create a parametric scattering route of optical transitions. Between the orbital ground states $\ket{1}$ and $\ket{2}$ a coherent excitation $B$ (spin wave) is created.}
	\label{fig4}
\end{figure}

In a final experiment, we reduced the strength of one of the laser fields, using it as a signal field, while the other, strong field, is used as a control field. We then directly measured the signal's gain and absorption induced by its interaction with the SiV\textsuperscript{-} ensemble driven by the control field. With the same blue detuning of $\Delta = 70$\,GHz as in the experiments above, here, in contrast to those experiments, the signal field is measured after transmission through the diamond sample. In order to measure the weak signal field, all other wavelengths are suppressed by a polarizer and a fibre-coupled grating filter setup \cite{supp}. Both absorption and amplification were observed for the weak signal field, depending on the relative phase between the signal and control arm of the setup. 
By fine-tuning the delay between the pulses near the point of full temporal overlap, the relative phase between them could be adjusted without significant change of the overlap itself. A maximum of \mbox{$\approx 80$\%} absorption and \mbox{$\approx 60$\%} amplification were reached after subtraction of the residual control field. All measured fields after filtering are shown in \mbox{Fig. \ref{fig4}\,(a)}. In measurements, saturation of absorption and amplification was reached when increasing the power of the control field \cite{supp}. Therefore, for the given detuning, both effects are most likely limited by the optical density of the sample.

Absorption and amplification can be explained by an off-resonant FWM process with an additional Stokes field. The signal and control field are generated in separate beam paths by etalons of the same type from the same laser pulse. As well as preparing the control field, this configuration produces unintentionally an additional field at the Stokes frequency by means of leakage through the low-energy tail of the control field etalon \cite{supp}. The Stokes, control, and signal pulses arrive at the same time after the pump pulse and with a fixed phase relation at the ensemble as can be seen in the schematic pulse scheme of Fig. \ref{fig4}\,(b). Within the SiV$^{-}$ ensemble, FWM takes place because the control field couples to both ground states, generating a double-$\Lambda$ system via two virtual excited states and, together with signal and Stokes fields, creates a parametric scattering route as indicated in \mbox{Fig. \ref{fig4}\,(c)}. Both $\Lambda$-systems interact via the ground state coherence $B$ (spin wave). The third-order susceptibility is enhanced by the parametric scattering route, increasing the FWM process in this thin ensemble of low optical density \cite{Harris1990}.

The interaction of the fields can be described by the linearised Maxwell-Bloch equations in the adiabatic approximation \cite{PhysRevA.96.012338}. These equations have been modified to account for the initial population distribution set by the limited pumping power and the rethermalization of the ground states \cite{supp}.

The relative phases of the signal, control, and Stokes fields determines whether the signal field will be absorbed or amplified \cite{supp}. The phase dependence arises from seeded Raman scattering where the phase of each field is set as an initial condition \cite{Belsley1993, Kosachiov1992}. Simulations show that the signal and Stokes fields are both amplified or absorbed under the same phase conditions. Therefore, an energy transfer between these two fields can be excluded and energy is only transferred between them and the control field. 

By applying active phase stabilization fixing the relative phase between the signal field and the arm containing control and Stokes field, we were able to switch between absorption and gain in a controlled way \cite{supp}.
Similar phase-sensitive effects involving resonant coherent population trapping in warm helium cells \cite{Lugani2016, Neveu}, as well as, off-resonant FWM interactions in a double-$\Lambda$ system in warm rubidium vapour \cite{Shahrokhshahi2018} have been observed recently.\\

In conclusion, we have shown both single-photon and two-photon coherent manipulation of an ensemble of SiV\textsuperscript{-} centers and demonstrated the potential of optical Hahn echo schemes to further remedy the effects of the residual inhomogeneous broadening. Moreover, we demonstrated strong light-matter interactions in this only 300\,nm thick ensemble by implementing STIRAP and by showing efficient absorption and amplification of weak coherent states by strong seeded off-resonant FWM, both depending on the relative phase between the signal, control, and Stokes pulses. This outlines the great potential of SiV\textsuperscript{-} ensembles to realize integrated devices relying on strong light-matter interactions and single photon nonlinearities. The ability to coherently manipulate ensembles of SiV$^{-}$ centers will enable Raman-based optical quantum memories \cite{nunn2007,gorshkov2007}. Furthermore, controlled parametric scattering routes reveal a wide applicability to single photon switches \cite{harris1998} or single photon nonlinearities \cite{liu2016,tiarks2016,beck2016}. The techniques employed in this work involving the orbital levels of the SiV$^{-}$ seamlessly extend to the center's spin sublevels, enabling access to a potentially long-lived degree of freedom \cite{PhysRevLett.120.053603, Sukachev2017}. Moreover, these experiments form the foundation for similar work using other inversion symmetric defects in diamond such as GeV \cite{Iwasaki2015}, SnV \cite{Iwasaki2017}, or NiV \cite{Larico2009} ensembles, which may combine outstanding spectral with optimized electron spin properties and even larger bandwidths due to enhanced spin-spin and spin-orbit interactions.\\

\begin{acknowledgments}
	The authors thank Benjamin Brecht, Patrick Ledingham, and Dylan Saunders for insightful discussions. This work was financially supported by the European Materials, Physical and Nanoscience (MPNS) Cooperation in Science \& Technology (COST) Action MP1403 (Nanoscale Quantum Optics) funding Short Term Scientific Missions (STSMs) of J.N.B. and C.W. C.W. acknowledges travel grants from Oriel College (Oxford). I.A.W. and C.W. were partially funded by an ERC Advanced Grant (MOQUACINO). J.N. was supported by an EPSRC Programme Grant (BLOQS), by the EPSRC Networked Quantum Information Technologies Hub (NQIT), and a Royal Society Fellowship. E.P. acknowledges a Marie Curie fellowship (PIEF-GA-2013-627372). This work has been partially funded by the European Community's Seventh Framework Programme (FP7/2007-2013) under Grant Agreement No. 611143 (DIADEMS) to C.B.\\
C.W., J.G., and J.N.B. contributed equally to this work.
\end{acknowledgments}

\bibliographystyle{apsrev4-1}

%


\begin{thebibliography}{55}%
\makeatletter
\providecommand \@ifxundefined [1]{%
 \@ifx{#1\undefined}
}%
\providecommand \@ifnum [1]{%
 \ifnum #1\expandafter \@firstoftwo
 \else \expandafter \@secondoftwo
 \fi
}%
\providecommand \@ifx [1]{%
 \ifx #1\expandafter \@firstoftwo
 \else \expandafter \@secondoftwo
 \fi
}%
\providecommand \natexlab [1]{#1}%
\providecommand \enquote  [1]{``#1''}%
\providecommand \bibnamefont  [1]{#1}%
\providecommand \bibfnamefont [1]{#1}%
\providecommand \citenamefont [1]{#1}%
\providecommand \href@noop [0]{\@secondoftwo}%
\providecommand \href [0]{\begingroup \@sanitize@url \@href}%
\providecommand \@href[1]{\@@startlink{#1}\@@href}%
\providecommand \@@href[1]{\endgroup#1\@@endlink}%
\providecommand \@sanitize@url [0]{\catcode `\\12\catcode `\$12\catcode
  `\&12\catcode `\#12\catcode `\^12\catcode `\_12\catcode `\%12\relax}%
\providecommand \@@startlink[1]{}%
\providecommand \@@endlink[0]{}%
\providecommand \url  [0]{\begingroup\@sanitize@url \@url }%
\providecommand \@url [1]{\endgroup\@href {#1}{\urlprefix }}%
\providecommand \urlprefix  [0]{URL }%
\providecommand \Eprint [0]{\href }%
\providecommand \doibase [0]{http://dx.doi.org/}%
\providecommand \selectlanguage [0]{\@gobble}%
\providecommand \bibinfo  [0]{\@secondoftwo}%
\providecommand \bibfield  [0]{\@secondoftwo}%
\providecommand \translation [1]{[#1]}%
\providecommand \BibitemOpen [0]{}%
\providecommand \bibitemStop [0]{}%
\providecommand \bibitemNoStop [0]{.\EOS\space}%
\providecommand \EOS [0]{\spacefactor3000\relax}%
\providecommand \BibitemShut  [1]{\csname bibitem#1\endcsname}%
\let\auto@bib@innerbib\@empty
\bibitem [{\citenamefont {Monroe}(2002)}]{Monroe2002}%
  \BibitemOpen
  \bibfield  {author} {\bibinfo {author} {\bibfnamefont {C.}~\bibnamefont
  {Monroe}},\ }\href {\doibase 10.1038/416238a} {\bibfield  {journal} {\bibinfo
   {journal} {Nature}\ }\textbf {\bibinfo {volume} {416}},\ \bibinfo {pages}
  {238} (\bibinfo {year} {2002})}\BibitemShut {NoStop}%
\bibitem [{\citenamefont {Kok}\ \emph {et~al.}(2007)\citenamefont {Kok},
  \citenamefont {Munro}, \citenamefont {Nemoto}, \citenamefont {Ralph},
  \citenamefont {Dowling},\ and\ \citenamefont {Milburn}}]{Kok2007}%
  \BibitemOpen
  \bibfield  {author} {\bibinfo {author} {\bibfnamefont {P.}~\bibnamefont
  {Kok}}, \bibinfo {author} {\bibfnamefont {W.~J.}\ \bibnamefont {Munro}},
  \bibinfo {author} {\bibfnamefont {K.}~\bibnamefont {Nemoto}}, \bibinfo
  {author} {\bibfnamefont {T.~C.}\ \bibnamefont {Ralph}}, \bibinfo {author}
  {\bibfnamefont {J.~P.}\ \bibnamefont {Dowling}}, \ and\ \bibinfo {author}
  {\bibfnamefont {G.~J.}\ \bibnamefont {Milburn}},\ }\href@noop {} {\bibfield
  {journal} {\bibinfo  {journal} {Reviews of Modern Physics}\ }\textbf
  {\bibinfo {volume} {79}},\ \bibinfo {pages} {135} (\bibinfo {year}
  {2007})}\BibitemShut {NoStop}%
\bibitem [{\citenamefont {Degen}\ \emph {et~al.}(2017)\citenamefont {Degen},
  \citenamefont {Reinhard},\ and\ \citenamefont {Cappellaro}}]{Degen2017}%
  \BibitemOpen
  \bibfield  {author} {\bibinfo {author} {\bibfnamefont {C.~L.}\ \bibnamefont
  {Degen}}, \bibinfo {author} {\bibfnamefont {F.}~\bibnamefont {Reinhard}}, \
  and\ \bibinfo {author} {\bibfnamefont {P.}~\bibnamefont {Cappellaro}},\
  }\href {\doibase 10.1103/RevModPhys.89.035002} {\bibfield  {journal}
  {\bibinfo  {journal} {Reviews of Modern Physics}\ }\textbf {\bibinfo {volume}
  {89}},\ \bibinfo {pages} {035002} (\bibinfo {year} {2017})}\BibitemShut
  {NoStop}%
\bibitem [{\citenamefont {Bernardi}\ \emph {et~al.}(2017)\citenamefont
  {Bernardi}, \citenamefont {Nelz}, \citenamefont {Sonusen},\ and\
  \citenamefont {Neu}}]{Bernardi2017}%
  \BibitemOpen
  \bibfield  {author} {\bibinfo {author} {\bibfnamefont {E.}~\bibnamefont
  {Bernardi}}, \bibinfo {author} {\bibfnamefont {R.}~\bibnamefont {Nelz}},
  \bibinfo {author} {\bibfnamefont {S.}~\bibnamefont {Sonusen}}, \ and\
  \bibinfo {author} {\bibfnamefont {E.}~\bibnamefont {Neu}},\ }\href
  {http://arxiv.org/abs/1704.04011 http://dx.doi.org/10.3390/cryst7050124}
  {\bibfield  {journal} {\bibinfo  {journal} {Crystals}\ }\textbf {\bibinfo
  {volume} {7}},\ \bibinfo {pages} {124} (\bibinfo {year} {2017})}\BibitemShut
  {NoStop}%
\bibitem [{\citenamefont {Doherty}\ \emph {et~al.}(2013)\citenamefont
  {Doherty}, \citenamefont {Manson}, \citenamefont {Delaney}, \citenamefont
  {Jelezko}, \citenamefont {Wrachtrup},\ and\ \citenamefont
  {Hollenberg}}]{Doherty2013}%
  \BibitemOpen
  \bibfield  {author} {\bibinfo {author} {\bibfnamefont {M.~W.}\ \bibnamefont
  {Doherty}}, \bibinfo {author} {\bibfnamefont {N.~B.}\ \bibnamefont {Manson}},
  \bibinfo {author} {\bibfnamefont {P.}~\bibnamefont {Delaney}}, \bibinfo
  {author} {\bibfnamefont {F.}~\bibnamefont {Jelezko}}, \bibinfo {author}
  {\bibfnamefont {J.}~\bibnamefont {Wrachtrup}}, \ and\ \bibinfo {author}
  {\bibfnamefont {L.~C.~L.}\ \bibnamefont {Hollenberg}},\ }\href {\doibase
  10.1016/j.physrep.2013.02.001} {\bibfield  {journal} {\bibinfo  {journal}
  {Physics Reports}\ }\textbf {\bibinfo {volume} {528}},\ \bibinfo {pages} {1}
  (\bibinfo {year} {2013})}\BibitemShut {NoStop}%
\bibitem [{\citenamefont {Aharonovich}\ \emph {et~al.}(2011)\citenamefont
  {Aharonovich}, \citenamefont {Castelletto}, \citenamefont {Simpson},
  \citenamefont {Su}, \citenamefont {Greentree},\ and\ \citenamefont
  {Prawer}}]{Aharonovich2011}%
  \BibitemOpen
  \bibfield  {author} {\bibinfo {author} {\bibfnamefont {I.}~\bibnamefont
  {Aharonovich}}, \bibinfo {author} {\bibfnamefont {S.}~\bibnamefont
  {Castelletto}}, \bibinfo {author} {\bibfnamefont {D.~A.}\ \bibnamefont
  {Simpson}}, \bibinfo {author} {\bibfnamefont {C.~H.}\ \bibnamefont {Su}},
  \bibinfo {author} {\bibfnamefont {A.~D.}\ \bibnamefont {Greentree}}, \ and\
  \bibinfo {author} {\bibfnamefont {S.}~\bibnamefont {Prawer}},\ }\href@noop {}
  {\bibfield  {journal} {\bibinfo  {journal} {Reports on Progress in Physics}\
  }\textbf {\bibinfo {volume} {74}},\ \bibinfo {pages} {076501} (\bibinfo
  {year} {2011})}\BibitemShut {NoStop}%
\bibitem [{\citenamefont {Hensen}\ \emph {et~al.}(2015)\citenamefont {Hensen},
  \citenamefont {Bernien}, \citenamefont {Dr\'{e}au}, \citenamefont {Reiserer},
  \citenamefont {Kalb}, \citenamefont {Blok}, \citenamefont {Ruitenberg},
  \citenamefont {Vermeulen}, \citenamefont {Schouten}, \citenamefont
  {Abell\'{a}n}, \citenamefont {Amaya}, \citenamefont {Pruneri}, \citenamefont
  {Mitchell}, \citenamefont {Markham}, \citenamefont {Twitchen}, \citenamefont
  {Elkouss}, \citenamefont {Wehner}, \citenamefont {Taminiau},\ and\
  \citenamefont {Hanson}}]{Hensen2015}%
  \BibitemOpen
  \bibfield  {author} {\bibinfo {author} {\bibfnamefont {B.}~\bibnamefont
  {Hensen}}, \bibinfo {author} {\bibfnamefont {H.}~\bibnamefont {Bernien}},
  \bibinfo {author} {\bibfnamefont {A.~E.}\ \bibnamefont {Dr\'{e}au}}, \bibinfo
  {author} {\bibfnamefont {A.}~\bibnamefont {Reiserer}}, \bibinfo {author}
  {\bibfnamefont {N.}~\bibnamefont {Kalb}}, \bibinfo {author} {\bibfnamefont
  {M.~S.}\ \bibnamefont {Blok}}, \bibinfo {author} {\bibfnamefont
  {J.}~\bibnamefont {Ruitenberg}}, \bibinfo {author} {\bibfnamefont {R.~F.~L.}\
  \bibnamefont {Vermeulen}}, \bibinfo {author} {\bibfnamefont {R.~N.}\
  \bibnamefont {Schouten}}, \bibinfo {author} {\bibfnamefont {C.}~\bibnamefont
  {Abell\'{a}n}}, \bibinfo {author} {\bibfnamefont {W.}~\bibnamefont {Amaya}},
  \bibinfo {author} {\bibfnamefont {V.}~\bibnamefont {Pruneri}}, \bibinfo
  {author} {\bibfnamefont {M.~W.}\ \bibnamefont {Mitchell}}, \bibinfo {author}
  {\bibfnamefont {M.}~\bibnamefont {Markham}}, \bibinfo {author} {\bibfnamefont
  {D.~J.}\ \bibnamefont {Twitchen}}, \bibinfo {author} {\bibfnamefont
  {D.}~\bibnamefont {Elkouss}}, \bibinfo {author} {\bibfnamefont
  {S.}~\bibnamefont {Wehner}}, \bibinfo {author} {\bibfnamefont {T.~H.}\
  \bibnamefont {Taminiau}}, \ and\ \bibinfo {author} {\bibfnamefont
  {R.}~\bibnamefont {Hanson}},\ }\href {\doibase 10.1038/nature15759}
  {\bibfield  {journal} {\bibinfo  {journal} {Nature}\ }\textbf {\bibinfo
  {volume} {526}},\ \bibinfo {pages} {682} (\bibinfo {year}
  {2015})}\BibitemShut {NoStop}%
\bibitem [{\citenamefont {Neu}\ \emph {et~al.}(2011)\citenamefont {Neu},
  \citenamefont {Steinmetz}, \citenamefont {Riedrich-M\"{o}ller}, \citenamefont
  {Gsell}, \citenamefont {Fischer}, \citenamefont {Schreck},\ and\
  \citenamefont {Becher}}]{Neu2011}%
  \BibitemOpen
  \bibfield  {author} {\bibinfo {author} {\bibfnamefont {E.}~\bibnamefont
  {Neu}}, \bibinfo {author} {\bibfnamefont {D.}~\bibnamefont {Steinmetz}},
  \bibinfo {author} {\bibfnamefont {J.}~\bibnamefont {Riedrich-M\"{o}ller}},
  \bibinfo {author} {\bibfnamefont {S.}~\bibnamefont {Gsell}}, \bibinfo
  {author} {\bibfnamefont {M.}~\bibnamefont {Fischer}}, \bibinfo {author}
  {\bibfnamefont {M.}~\bibnamefont {Schreck}}, \ and\ \bibinfo {author}
  {\bibfnamefont {C.}~\bibnamefont {Becher}},\ }\href@noop {} {\bibfield
  {journal} {\bibinfo  {journal} {New Journal of Physics}\ }\textbf {\bibinfo
  {volume} {13}},\ \bibinfo {pages} {025012} (\bibinfo {year}
  {2011})}\BibitemShut {NoStop}%
\bibitem [{\citenamefont {Sipahigil}\ \emph {et~al.}(2014)\citenamefont
  {Sipahigil}, \citenamefont {Jahnke}, \citenamefont {Rogers}, \citenamefont
  {Teraji}, \citenamefont {Isoya}, \citenamefont {Zibrov}, \citenamefont
  {Jelezko},\ and\ \citenamefont {Lukin}}]{Sipahigil2014}%
  \BibitemOpen
  \bibfield  {author} {\bibinfo {author} {\bibfnamefont {A.}~\bibnamefont
  {Sipahigil}}, \bibinfo {author} {\bibfnamefont {K.~D.}\ \bibnamefont
  {Jahnke}}, \bibinfo {author} {\bibfnamefont {L.~J.}\ \bibnamefont {Rogers}},
  \bibinfo {author} {\bibfnamefont {T.}~\bibnamefont {Teraji}}, \bibinfo
  {author} {\bibfnamefont {J.}~\bibnamefont {Isoya}}, \bibinfo {author}
  {\bibfnamefont {A.~S.}\ \bibnamefont {Zibrov}}, \bibinfo {author}
  {\bibfnamefont {F.}~\bibnamefont {Jelezko}}, \ and\ \bibinfo {author}
  {\bibfnamefont {M.~D.}\ \bibnamefont {Lukin}},\ }\href {\doibase
  10.1103/PhysRevLett.113.113602} {\bibfield  {journal} {\bibinfo  {journal}
  {Physical Review Letters}\ }\textbf {\bibinfo {volume} {113}},\ \bibinfo
  {pages} {113602} (\bibinfo {year} {2014})}\BibitemShut {NoStop}%
\bibitem [{\citenamefont {Riedrich-M\"{o}ller}\ \emph
  {et~al.}(2014)\citenamefont {Riedrich-M\"{o}ller}, \citenamefont {Arend},
  \citenamefont {Pauly}, \citenamefont {M\"{u}cklich}, \citenamefont {Fischer},
  \citenamefont {Gsell}, \citenamefont {Schreck},\ and\ \citenamefont
  {Becher}}]{Riedrich-m2012}%
  \BibitemOpen
  \bibfield  {author} {\bibinfo {author} {\bibfnamefont {J.}~\bibnamefont
  {Riedrich-M\"{o}ller}}, \bibinfo {author} {\bibfnamefont {C.}~\bibnamefont
  {Arend}}, \bibinfo {author} {\bibfnamefont {C.}~\bibnamefont {Pauly}},
  \bibinfo {author} {\bibfnamefont {F.}~\bibnamefont {M\"{u}cklich}}, \bibinfo
  {author} {\bibfnamefont {M.}~\bibnamefont {Fischer}}, \bibinfo {author}
  {\bibfnamefont {S.}~\bibnamefont {Gsell}}, \bibinfo {author} {\bibfnamefont
  {M.}~\bibnamefont {Schreck}}, \ and\ \bibinfo {author} {\bibfnamefont
  {C.}~\bibnamefont {Becher}},\ }\href {\doibase 10.1021/nl502327b} {\bibfield
  {journal} {\bibinfo  {journal} {Nano Letters}\ }\textbf {\bibinfo {volume}
  {14}},\ \bibinfo {pages} {5281} (\bibinfo {year} {2014})}\BibitemShut
  {NoStop}%
\bibitem [{\citenamefont {Sipahigil}\ \emph {et~al.}(2016)\citenamefont
  {Sipahigil}, \citenamefont {Evans}, \citenamefont {Sukachev}, \citenamefont
  {Burek}, \citenamefont {Borregaard}, \citenamefont {Bhaskar}, \citenamefont
  {Nguyen}, \citenamefont {Pacheco}, \citenamefont {Atikian}, \citenamefont
  {Meuwly}, \citenamefont {Camacho}, \citenamefont {Jelezko}, \citenamefont
  {Bielejec}, \citenamefont {Park}, \citenamefont {Lon\v{c}ar},\ and\
  \citenamefont {Lukin}}]{Sipahigil2016}%
  \BibitemOpen
  \bibfield  {author} {\bibinfo {author} {\bibfnamefont {A.}~\bibnamefont
  {Sipahigil}}, \bibinfo {author} {\bibfnamefont {R.~E.}\ \bibnamefont
  {Evans}}, \bibinfo {author} {\bibfnamefont {D.~D.}\ \bibnamefont {Sukachev}},
  \bibinfo {author} {\bibfnamefont {M.~J.}\ \bibnamefont {Burek}}, \bibinfo
  {author} {\bibfnamefont {J.}~\bibnamefont {Borregaard}}, \bibinfo {author}
  {\bibfnamefont {M.~K.}\ \bibnamefont {Bhaskar}}, \bibinfo {author}
  {\bibfnamefont {C.~T.}\ \bibnamefont {Nguyen}}, \bibinfo {author}
  {\bibfnamefont {J.~L.}\ \bibnamefont {Pacheco}}, \bibinfo {author}
  {\bibfnamefont {H.~A.}\ \bibnamefont {Atikian}}, \bibinfo {author}
  {\bibfnamefont {C.}~\bibnamefont {Meuwly}}, \bibinfo {author} {\bibfnamefont
  {R.~M.}\ \bibnamefont {Camacho}}, \bibinfo {author} {\bibfnamefont
  {F.}~\bibnamefont {Jelezko}}, \bibinfo {author} {\bibfnamefont
  {E.}~\bibnamefont {Bielejec}}, \bibinfo {author} {\bibfnamefont
  {H.}~\bibnamefont {Park}}, \bibinfo {author} {\bibfnamefont {M.}~\bibnamefont
  {Lon\v{c}ar}}, \ and\ \bibinfo {author} {\bibfnamefont {M.~D.}\ \bibnamefont
  {Lukin}},\ }\href {\doibase 10.1126/science.aah6875} {\bibfield  {journal}
  {\bibinfo  {journal} {Science}\ }\textbf {\bibinfo {volume} {354}},\ \bibinfo
  {pages} {847} (\bibinfo {year} {2016})}\BibitemShut {NoStop}%
\bibitem [{\citenamefont {Pingault}\ \emph {et~al.}(2017)\citenamefont
  {Pingault}, \citenamefont {Jarausch}, \citenamefont {Hepp}, \citenamefont
  {Klintberg}, \citenamefont {Becker}, \citenamefont {Markham}, \citenamefont
  {Becher},\ and\ \citenamefont {Atat\"{u}re}}]{Pingault2017}%
  \BibitemOpen
  \bibfield  {author} {\bibinfo {author} {\bibfnamefont {B.}~\bibnamefont
  {Pingault}}, \bibinfo {author} {\bibfnamefont {D.-D.}\ \bibnamefont
  {Jarausch}}, \bibinfo {author} {\bibfnamefont {C.}~\bibnamefont {Hepp}},
  \bibinfo {author} {\bibfnamefont {L.}~\bibnamefont {Klintberg}}, \bibinfo
  {author} {\bibfnamefont {J.~N.}\ \bibnamefont {Becker}}, \bibinfo {author}
  {\bibfnamefont {M.}~\bibnamefont {Markham}}, \bibinfo {author} {\bibfnamefont
  {C.}~\bibnamefont {Becher}}, \ and\ \bibinfo {author} {\bibfnamefont
  {M.}~\bibnamefont {Atat\"{u}re}},\ }\href {\doibase 10.1038/ncomms15579}
  {\bibfield  {journal} {\bibinfo  {journal} {Nature Communications}\ }\textbf
  {\bibinfo {volume} {8}},\ \bibinfo {pages} {15579} (\bibinfo {year}
  {2017})}\BibitemShut {NoStop}%
\bibitem [{\citenamefont {Becker}\ \emph {et~al.}(2016)\citenamefont {Becker},
  \citenamefont {G\"{o}rlitz}, \citenamefont {Arend}, \citenamefont {Markham},\
  and\ \citenamefont {Becher}}]{Becker2016}%
  \BibitemOpen
  \bibfield  {author} {\bibinfo {author} {\bibfnamefont {J.~N.}\ \bibnamefont
  {Becker}}, \bibinfo {author} {\bibfnamefont {J.}~\bibnamefont {G\"{o}rlitz}},
  \bibinfo {author} {\bibfnamefont {C.}~\bibnamefont {Arend}}, \bibinfo
  {author} {\bibfnamefont {M.}~\bibnamefont {Markham}}, \ and\ \bibinfo
  {author} {\bibfnamefont {C.}~\bibnamefont {Becher}},\ }\href
  {http://www.nature.com/doifinder/10.1038/ncomms13512} {\bibfield  {journal}
  {\bibinfo  {journal} {Nature Communications}\ }\textbf {\bibinfo {volume}
  {7}},\ \bibinfo {pages} {13512} (\bibinfo {year} {2016})}\BibitemShut
  {NoStop}%
\bibitem [{\citenamefont {Becker}\ \emph {et~al.}(2018)\citenamefont {Becker},
  \citenamefont {Pingault}, \citenamefont {Gro\ss{}}, \citenamefont
  {G\"undo\ifmmode~\breve{g}\else \u{g}\fi{}an}, \citenamefont {Kukharchyk},
  \citenamefont {Markham}, \citenamefont {Edmonds}, \citenamefont {Atat\"ure},
  \citenamefont {Bushev},\ and\ \citenamefont
  {Becher}}]{PhysRevLett.120.053603}%
  \BibitemOpen
  \bibfield  {author} {\bibinfo {author} {\bibfnamefont {J.~N.}\ \bibnamefont
  {Becker}}, \bibinfo {author} {\bibfnamefont {B.}~\bibnamefont {Pingault}},
  \bibinfo {author} {\bibfnamefont {D.}~\bibnamefont {Gro\ss{}}}, \bibinfo
  {author} {\bibfnamefont {M.}~\bibnamefont {G\"undo\ifmmode~\breve{g}\else
  \u{g}\fi{}an}}, \bibinfo {author} {\bibfnamefont {N.}~\bibnamefont
  {Kukharchyk}}, \bibinfo {author} {\bibfnamefont {M.}~\bibnamefont {Markham}},
  \bibinfo {author} {\bibfnamefont {A.}~\bibnamefont {Edmonds}}, \bibinfo
  {author} {\bibfnamefont {M.}~\bibnamefont {Atat\"ure}}, \bibinfo {author}
  {\bibfnamefont {P.}~\bibnamefont {Bushev}}, \ and\ \bibinfo {author}
  {\bibfnamefont {C.}~\bibnamefont {Becher}},\ }\href {\doibase
  10.1103/PhysRevLett.120.053603} {\bibfield  {journal} {\bibinfo  {journal}
  {Phys. Rev. Lett.}\ }\textbf {\bibinfo {volume} {120}},\ \bibinfo {pages}
  {053603} (\bibinfo {year} {2018})}\BibitemShut {NoStop}%
\bibitem [{\citenamefont {Togan}\ \emph {et~al.}(2010)\citenamefont {Togan},
  \citenamefont {Chu}, \citenamefont {Trifonov}, \citenamefont {Jiang},
  \citenamefont {Maze}, \citenamefont {Childress}, \citenamefont {Dutt},
  \citenamefont {S\o{}rensen}, \citenamefont {Hemmer}, \citenamefont {Zibrov},\
  and\ \citenamefont {Lukin}}]{Togan2010}%
  \BibitemOpen
  \bibfield  {author} {\bibinfo {author} {\bibfnamefont {E.}~\bibnamefont
  {Togan}}, \bibinfo {author} {\bibfnamefont {Y.}~\bibnamefont {Chu}}, \bibinfo
  {author} {\bibfnamefont {A.~S.}\ \bibnamefont {Trifonov}}, \bibinfo {author}
  {\bibfnamefont {L.}~\bibnamefont {Jiang}}, \bibinfo {author} {\bibfnamefont
  {J.}~\bibnamefont {Maze}}, \bibinfo {author} {\bibfnamefont {L.}~\bibnamefont
  {Childress}}, \bibinfo {author} {\bibfnamefont {M.~V.~G.}\ \bibnamefont
  {Dutt}}, \bibinfo {author} {\bibfnamefont {A.~S.}\ \bibnamefont
  {S\o{}rensen}}, \bibinfo {author} {\bibfnamefont {P.~R.}\ \bibnamefont
  {Hemmer}}, \bibinfo {author} {\bibfnamefont {A.~S.}\ \bibnamefont {Zibrov}},
  \ and\ \bibinfo {author} {\bibfnamefont {M.~D.}\ \bibnamefont {Lukin}},\
  }\href {\doibase 10.1038/nature09256} {\bibfield  {journal} {\bibinfo
  {journal} {Nature}\ }\textbf {\bibinfo {volume} {466}},\ \bibinfo {pages}
  {730} (\bibinfo {year} {2010})}\BibitemShut {NoStop}%
\bibitem [{\citenamefont {Arend}\ \emph {et~al.}(2016)\citenamefont {Arend},
  \citenamefont {Becker}, \citenamefont {Sternschulte}, \citenamefont
  {Steinm\"{u}ller-Nethl},\ and\ \citenamefont {Becher}}]{Arend2016}%
  \BibitemOpen
  \bibfield  {author} {\bibinfo {author} {\bibfnamefont {C.}~\bibnamefont
  {Arend}}, \bibinfo {author} {\bibfnamefont {J.~N.}\ \bibnamefont {Becker}},
  \bibinfo {author} {\bibfnamefont {H.}~\bibnamefont {Sternschulte}}, \bibinfo
  {author} {\bibfnamefont {D.}~\bibnamefont {Steinm\"{u}ller-Nethl}}, \ and\
  \bibinfo {author} {\bibfnamefont {C.}~\bibnamefont {Becher}},\ }\href@noop {}
  {\bibfield  {journal} {\bibinfo  {journal} {Physical Review B - Condensed
  Matter and Materials Physics}\ }\textbf {\bibinfo {volume} {94}},\ \bibinfo
  {pages} {045203} (\bibinfo {year} {2016})}\BibitemShut {NoStop}%
\bibitem [{\citenamefont {Hepp}\ \emph {et~al.}(2014)\citenamefont {Hepp},
  \citenamefont {M\"{u}ller}, \citenamefont {Waselowski}, \citenamefont
  {Becker}, \citenamefont {Pingault}, \citenamefont {Sternschulte},
  \citenamefont {Steinm\"{u}ller-Nethl}, \citenamefont {Gali}, \citenamefont
  {Maze}, \citenamefont {Atat\"{u}re},\ and\ \citenamefont
  {Becher}}]{Hepp2014}%
  \BibitemOpen
  \bibfield  {author} {\bibinfo {author} {\bibfnamefont {C.}~\bibnamefont
  {Hepp}}, \bibinfo {author} {\bibfnamefont {T.}~\bibnamefont {M\"{u}ller}},
  \bibinfo {author} {\bibfnamefont {V.}~\bibnamefont {Waselowski}}, \bibinfo
  {author} {\bibfnamefont {J.~N.}\ \bibnamefont {Becker}}, \bibinfo {author}
  {\bibfnamefont {B.}~\bibnamefont {Pingault}}, \bibinfo {author}
  {\bibfnamefont {H.}~\bibnamefont {Sternschulte}}, \bibinfo {author}
  {\bibfnamefont {D.}~\bibnamefont {Steinm\"{u}ller-Nethl}}, \bibinfo {author}
  {\bibfnamefont {A.}~\bibnamefont {Gali}}, \bibinfo {author} {\bibfnamefont
  {J.~R.}\ \bibnamefont {Maze}}, \bibinfo {author} {\bibfnamefont
  {M.}~\bibnamefont {Atat\"{u}re}}, \ and\ \bibinfo {author} {\bibfnamefont
  {C.}~\bibnamefont {Becher}},\ }\href@noop {} {\bibfield  {journal} {\bibinfo
  {journal} {Physical Review Letters}\ }\textbf {\bibinfo {volume} {112}},\
  \bibinfo {pages} {036405} (\bibinfo {year} {2014})}\BibitemShut {NoStop}%
\bibitem [{\citenamefont {Gali}\ and\ \citenamefont {Maze}(2013)}]{gali2013}%
  \BibitemOpen
  \bibfield  {author} {\bibinfo {author} {\bibfnamefont {A.}~\bibnamefont
  {Gali}}\ and\ \bibinfo {author} {\bibfnamefont {J.~R.}\ \bibnamefont
  {Maze}},\ }\href {https://link.aps.org/doi/10.1103/PhysRevB.88.235205}
  {\bibfield  {journal} {\bibinfo  {journal} {Phys. Rev. B}\ }\textbf {\bibinfo
  {volume} {88}},\ \bibinfo {pages} {235205} (\bibinfo {year}
  {2013})}\BibitemShut {NoStop}%
\bibitem [{\citenamefont {Pingault}\ \emph {et~al.}(2014)\citenamefont
  {Pingault}, \citenamefont {Becker}, \citenamefont {Schulte}, \citenamefont
  {Arend}, \citenamefont {Hepp}, \citenamefont {Godde}, \citenamefont
  {Tartakovskii}, \citenamefont {Markham}, \citenamefont {Becher},\ and\
  \citenamefont {Atat\"ure}}]{Pingault2014}%
  \BibitemOpen
  \bibfield  {author} {\bibinfo {author} {\bibfnamefont {B.}~\bibnamefont
  {Pingault}}, \bibinfo {author} {\bibfnamefont {J.~N.}\ \bibnamefont
  {Becker}}, \bibinfo {author} {\bibfnamefont {C.~H.~H.}\ \bibnamefont
  {Schulte}}, \bibinfo {author} {\bibfnamefont {C.}~\bibnamefont {Arend}},
  \bibinfo {author} {\bibfnamefont {C.}~\bibnamefont {Hepp}}, \bibinfo {author}
  {\bibfnamefont {T.}~\bibnamefont {Godde}}, \bibinfo {author} {\bibfnamefont
  {A.~I.}\ \bibnamefont {Tartakovskii}}, \bibinfo {author} {\bibfnamefont
  {M.}~\bibnamefont {Markham}}, \bibinfo {author} {\bibfnamefont
  {C.}~\bibnamefont {Becher}}, \ and\ \bibinfo {author} {\bibfnamefont
  {M.}~\bibnamefont {Atat\"ure}},\ }\href {\doibase
  10.1103/PhysRevLett.113.263601} {\bibfield  {journal} {\bibinfo  {journal}
  {Phys. Rev. Lett.}\ }\textbf {\bibinfo {volume} {113}},\ \bibinfo {pages}
  {263601} (\bibinfo {year} {2014})}\BibitemShut {NoStop}%
\bibitem [{\citenamefont {Rogers}\ \emph {et~al.}(2014)\citenamefont {Rogers},
  \citenamefont {Jahnke}, \citenamefont {Metsch}, \citenamefont {Sipahigil},
  \citenamefont {Binder}, \citenamefont {Teraji}, \citenamefont {Sumiya},
  \citenamefont {Isoya}, \citenamefont {Lukin}, \citenamefont {Hemmer},\ and\
  \citenamefont {Jelezko}}]{Rogers20142}%
  \BibitemOpen
  \bibfield  {author} {\bibinfo {author} {\bibfnamefont {L.~J.}\ \bibnamefont
  {Rogers}}, \bibinfo {author} {\bibfnamefont {K.~D.}\ \bibnamefont {Jahnke}},
  \bibinfo {author} {\bibfnamefont {M.~H.}\ \bibnamefont {Metsch}}, \bibinfo
  {author} {\bibfnamefont {A.}~\bibnamefont {Sipahigil}}, \bibinfo {author}
  {\bibfnamefont {J.~M.}\ \bibnamefont {Binder}}, \bibinfo {author}
  {\bibfnamefont {T.}~\bibnamefont {Teraji}}, \bibinfo {author} {\bibfnamefont
  {H.}~\bibnamefont {Sumiya}}, \bibinfo {author} {\bibfnamefont
  {J.}~\bibnamefont {Isoya}}, \bibinfo {author} {\bibfnamefont {M.~D.}\
  \bibnamefont {Lukin}}, \bibinfo {author} {\bibfnamefont {P.}~\bibnamefont
  {Hemmer}}, \ and\ \bibinfo {author} {\bibfnamefont {F.}~\bibnamefont
  {Jelezko}},\ }\href {\doibase 10.1103/PhysRevLett.113.263602} {\bibfield
  {journal} {\bibinfo  {journal} {Phys. Rev. Lett.}\ }\textbf {\bibinfo
  {volume} {113}},\ \bibinfo {pages} {263602} (\bibinfo {year}
  {2014})}\BibitemShut {NoStop}%
\bibitem [{\citenamefont {Jahnke}\ \emph {et~al.}(2015)\citenamefont {Jahnke},
  \citenamefont {Sipahigil}, \citenamefont {Binder}, \citenamefont {Doherty},
  \citenamefont {Metsch}, \citenamefont {Rogers}, \citenamefont {Manson},
  \citenamefont {Lukin},\ and\ \citenamefont {Jelezko}}]{Jahnke2015}%
  \BibitemOpen
  \bibfield  {author} {\bibinfo {author} {\bibfnamefont {K.~D.}\ \bibnamefont
  {Jahnke}}, \bibinfo {author} {\bibfnamefont {A.}~\bibnamefont {Sipahigil}},
  \bibinfo {author} {\bibfnamefont {J.~M.}\ \bibnamefont {Binder}}, \bibinfo
  {author} {\bibfnamefont {M.~W.}\ \bibnamefont {Doherty}}, \bibinfo {author}
  {\bibfnamefont {M.}~\bibnamefont {Metsch}}, \bibinfo {author} {\bibfnamefont
  {L.~J.}\ \bibnamefont {Rogers}}, \bibinfo {author} {\bibfnamefont {N.~B.}\
  \bibnamefont {Manson}}, \bibinfo {author} {\bibfnamefont {M.~D.}\
  \bibnamefont {Lukin}}, \ and\ \bibinfo {author} {\bibfnamefont
  {F.}~\bibnamefont {Jelezko}},\ }\href
  {http://stacks.iop.org/1367-2630/17/i=4/a=043011} {\bibfield  {journal}
  {\bibinfo  {journal} {New Journal of Physics}\ }\textbf {\bibinfo {volume}
  {17}},\ \bibinfo {pages} {043011} (\bibinfo {year} {2015})}\BibitemShut
  {NoStop}%
\bibitem [{\citenamefont {Sukachev}\ \emph {et~al.}(2017)\citenamefont
  {Sukachev}, \citenamefont {Sipahigil}, \citenamefont {Nguyen}, \citenamefont
  {Bhaskar}, \citenamefont {Evans}, \citenamefont {Jelezko},\ and\
  \citenamefont {Lukin}}]{Sukachev2017}%
  \BibitemOpen
  \bibfield  {author} {\bibinfo {author} {\bibfnamefont {D.~D.}\ \bibnamefont
  {Sukachev}}, \bibinfo {author} {\bibfnamefont {A.}~\bibnamefont {Sipahigil}},
  \bibinfo {author} {\bibfnamefont {C.~T.}\ \bibnamefont {Nguyen}}, \bibinfo
  {author} {\bibfnamefont {M.~K.}\ \bibnamefont {Bhaskar}}, \bibinfo {author}
  {\bibfnamefont {R.~E.}\ \bibnamefont {Evans}}, \bibinfo {author}
  {\bibfnamefont {F.}~\bibnamefont {Jelezko}}, \ and\ \bibinfo {author}
  {\bibfnamefont {M.~D.}\ \bibnamefont {Lukin}},\ }\href {\doibase
  10.1103/PhysRevLett.119.223602} {\bibfield  {journal} {\bibinfo  {journal}
  {Physical Review Letters}\ }\textbf {\bibinfo {volume} {119}},\ \bibinfo
  {pages} {223602} (\bibinfo {year} {2017})}\BibitemShut {NoStop}%
\bibitem [{\citenamefont {Neu}\ \emph {et~al.}(2013)\citenamefont {Neu},
  \citenamefont {Hepp}, \citenamefont {Hauschild}, \citenamefont {Gsell},
  \citenamefont {Fischer}, \citenamefont {Sternschulte}, \citenamefont
  {Steinm\"{u}ller-Nethl}, \citenamefont {Schreck},\ and\ \citenamefont
  {Becher}}]{Neu2013}%
  \BibitemOpen
  \bibfield  {author} {\bibinfo {author} {\bibfnamefont {E.}~\bibnamefont
  {Neu}}, \bibinfo {author} {\bibfnamefont {C.}~\bibnamefont {Hepp}}, \bibinfo
  {author} {\bibfnamefont {M.}~\bibnamefont {Hauschild}}, \bibinfo {author}
  {\bibfnamefont {S.}~\bibnamefont {Gsell}}, \bibinfo {author} {\bibfnamefont
  {M.}~\bibnamefont {Fischer}}, \bibinfo {author} {\bibfnamefont
  {H.}~\bibnamefont {Sternschulte}}, \bibinfo {author} {\bibfnamefont
  {D.}~\bibnamefont {Steinm\"{u}ller-Nethl}}, \bibinfo {author} {\bibfnamefont
  {M.}~\bibnamefont {Schreck}}, \ and\ \bibinfo {author} {\bibfnamefont
  {C.}~\bibnamefont {Becher}},\ }\href@noop {} {\bibfield  {journal} {\bibinfo
  {journal} {New Journal of Physics}\ }\textbf {\bibinfo {volume} {15}},\
  \bibinfo {pages} {043005} (\bibinfo {year} {2013})}\BibitemShut {NoStop}%
\bibitem [{\citenamefont {Alzetta}\ \emph {et~al.}(1976)\citenamefont
  {Alzetta}, \citenamefont {Gozzini}, \citenamefont {Moi},\ and\ \citenamefont
  {Orriols}}]{Alzetta1976}%
  \BibitemOpen
  \bibfield  {author} {\bibinfo {author} {\bibfnamefont {G.}~\bibnamefont
  {Alzetta}}, \bibinfo {author} {\bibfnamefont {A.}~\bibnamefont {Gozzini}},
  \bibinfo {author} {\bibfnamefont {L.}~\bibnamefont {Moi}}, \ and\ \bibinfo
  {author} {\bibfnamefont {G.}~\bibnamefont {Orriols}},\ }\href {\doibase
  10.1007/BF02749417} {\bibfield  {journal} {\bibinfo  {journal} {Il Nuovo
  Cimento B (1971-1996)}\ }\textbf {\bibinfo {volume} {36}},\ \bibinfo {pages}
  {5} (\bibinfo {year} {1976})}\BibitemShut {NoStop}%
\bibitem [{\citenamefont {Gaubatz}\ \emph {et~al.}(1990)\citenamefont
  {Gaubatz}, \citenamefont {Rudecki}, \citenamefont {Schiemann},\ and\
  \citenamefont {Bergmann}}]{Gaubatz1990}%
  \BibitemOpen
  \bibfield  {author} {\bibinfo {author} {\bibfnamefont {U.}~\bibnamefont
  {Gaubatz}}, \bibinfo {author} {\bibfnamefont {P.}~\bibnamefont {Rudecki}},
  \bibinfo {author} {\bibfnamefont {S.}~\bibnamefont {Schiemann}}, \ and\
  \bibinfo {author} {\bibfnamefont {K.}~\bibnamefont {Bergmann}},\ }\href
  {\doibase 10.1063/1.458514} {\bibfield  {journal} {\bibinfo  {journal} {The
  Journal of Chemical Physics}\ }\textbf {\bibinfo {volume} {92}},\ \bibinfo
  {pages} {5363} (\bibinfo {year} {1990})}\BibitemShut {NoStop}%
\bibitem [{\citenamefont {Nunn}\ \emph {et~al.}(2007)\citenamefont {Nunn},
  \citenamefont {Walmsley}, \citenamefont {Raymer}, \citenamefont {Surmacz},
  \citenamefont {Waldermann}, \citenamefont {Wang},\ and\ \citenamefont
  {Jaksch}}]{nunn2007}%
  \BibitemOpen
  \bibfield  {author} {\bibinfo {author} {\bibfnamefont {J.}~\bibnamefont
  {Nunn}}, \bibinfo {author} {\bibfnamefont {I.~A.}\ \bibnamefont {Walmsley}},
  \bibinfo {author} {\bibfnamefont {M.~G.}\ \bibnamefont {Raymer}}, \bibinfo
  {author} {\bibfnamefont {K.}~\bibnamefont {Surmacz}}, \bibinfo {author}
  {\bibfnamefont {F.~C.}\ \bibnamefont {Waldermann}}, \bibinfo {author}
  {\bibfnamefont {Z.}~\bibnamefont {Wang}}, \ and\ \bibinfo {author}
  {\bibfnamefont {D.}~\bibnamefont {Jaksch}},\ }\href {\doibase
  10.1103/PhysRevA.75.011401} {\bibfield  {journal} {\bibinfo  {journal} {Phys.
  Rev. A}\ }\textbf {\bibinfo {volume} {75}},\ \bibinfo {pages} {011401(R)}
  (\bibinfo {year} {2007})}\BibitemShut {NoStop}%
\bibitem [{\citenamefont {Goss}\ \emph {et~al.}(1996)\citenamefont {Goss},
  \citenamefont {Jones}, \citenamefont {Breuer}, \citenamefont {Briddon},\ and\
  \citenamefont {\"{O}berg}}]{Goss1996}%
  \BibitemOpen
  \bibfield  {author} {\bibinfo {author} {\bibfnamefont {J.}~\bibnamefont
  {Goss}}, \bibinfo {author} {\bibfnamefont {R.}~\bibnamefont {Jones}},
  \bibinfo {author} {\bibfnamefont {S.}~\bibnamefont {Breuer}}, \bibinfo
  {author} {\bibfnamefont {P.}~\bibnamefont {Briddon}}, \ and\ \bibinfo
  {author} {\bibfnamefont {S.}~\bibnamefont {\"{O}berg}},\ }\href {\doibase
  10.1103/PhysRevLett.77.3041} {\bibfield  {journal} {\bibinfo  {journal}
  {Physical Review Letters}\ }\textbf {\bibinfo {volume} {77}},\ \bibinfo
  {pages} {3041} (\bibinfo {year} {1996})}\BibitemShut {NoStop}%
\bibitem [{\citenamefont {Sternschulte}\ \emph {et~al.}(1994)\citenamefont
  {Sternschulte}, \citenamefont {Thonke}, \citenamefont {Sauer}, \citenamefont
  {M\"{u}nzinger},\ and\ \citenamefont {Michler}}]{Sternschulte1994}%
  \BibitemOpen
  \bibfield  {author} {\bibinfo {author} {\bibfnamefont {H.}~\bibnamefont
  {Sternschulte}}, \bibinfo {author} {\bibfnamefont {K.}~\bibnamefont
  {Thonke}}, \bibinfo {author} {\bibfnamefont {R.}~\bibnamefont {Sauer}},
  \bibinfo {author} {\bibfnamefont {P.~C.}\ \bibnamefont {M\"{u}nzinger}}, \
  and\ \bibinfo {author} {\bibfnamefont {P.}~\bibnamefont {Michler}},\ }\href
  {\doibase 10.1103/PhysRevB.50.14554} {\bibfield  {journal} {\bibinfo
  {journal} {Physical Review B}\ }\textbf {\bibinfo {volume} {50}},\ \bibinfo
  {pages} {14554} (\bibinfo {year} {1994})}\BibitemShut {NoStop}%
\bibitem [{\citenamefont {Acosta}\ \emph {et~al.}(2009)\citenamefont {Acosta},
  \citenamefont {Bauch}, \citenamefont {Ledbetter}, \citenamefont {Santori},
  \citenamefont {Fu}, \citenamefont {Barclay}, \citenamefont {Beausoleil},
  \citenamefont {Linget}, \citenamefont {Roch}, \citenamefont {Treussart},
  \citenamefont {Chemerisov}, \citenamefont {Gawlik},\ and\ \citenamefont
  {Budker}}]{Acosta2009}%
  \BibitemOpen
  \bibfield  {author} {\bibinfo {author} {\bibfnamefont {V.~M.}\ \bibnamefont
  {Acosta}}, \bibinfo {author} {\bibfnamefont {E.}~\bibnamefont {Bauch}},
  \bibinfo {author} {\bibfnamefont {M.~P.}\ \bibnamefont {Ledbetter}}, \bibinfo
  {author} {\bibfnamefont {C.}~\bibnamefont {Santori}}, \bibinfo {author}
  {\bibfnamefont {K.~M.}\ \bibnamefont {Fu}}, \bibinfo {author} {\bibfnamefont
  {P.~E.}\ \bibnamefont {Barclay}}, \bibinfo {author} {\bibfnamefont {R.~G.}\
  \bibnamefont {Beausoleil}}, \bibinfo {author} {\bibfnamefont
  {H.}~\bibnamefont {Linget}}, \bibinfo {author} {\bibfnamefont {J.~F.}\
  \bibnamefont {Roch}}, \bibinfo {author} {\bibfnamefont {F.}~\bibnamefont
  {Treussart}}, \bibinfo {author} {\bibfnamefont {S.}~\bibnamefont
  {Chemerisov}}, \bibinfo {author} {\bibfnamefont {W.}~\bibnamefont {Gawlik}},
  \ and\ \bibinfo {author} {\bibfnamefont {D.}~\bibnamefont {Budker}},\
  }\href@noop {} {\bibfield  {journal} {\bibinfo  {journal} {Physical Review B
  - Condensed Matter and Materials Physics}\ }\textbf {\bibinfo {volume}
  {80}},\ \bibinfo {pages} {115202} (\bibinfo {year} {2009})}\BibitemShut
  {NoStop}%
\bibitem [{\citenamefont {Poem}\ \emph {et~al.}(2015)\citenamefont {Poem},
  \citenamefont {Weinzetl}, \citenamefont {Klatzow}, \citenamefont {Kaczmarek},
  \citenamefont {Munns}, \citenamefont {Champion}, \citenamefont {Saunders},
  \citenamefont {Nunn},\ and\ \citenamefont {Walmsley}}]{Poem2015}%
  \BibitemOpen
  \bibfield  {author} {\bibinfo {author} {\bibfnamefont {E.}~\bibnamefont
  {Poem}}, \bibinfo {author} {\bibfnamefont {C.}~\bibnamefont {Weinzetl}},
  \bibinfo {author} {\bibfnamefont {J.}~\bibnamefont {Klatzow}}, \bibinfo
  {author} {\bibfnamefont {K.~T.}\ \bibnamefont {Kaczmarek}}, \bibinfo {author}
  {\bibfnamefont {J.~H.~D.}\ \bibnamefont {Munns}}, \bibinfo {author}
  {\bibfnamefont {T.~F.~M.}\ \bibnamefont {Champion}}, \bibinfo {author}
  {\bibfnamefont {D.~J.}\ \bibnamefont {Saunders}}, \bibinfo {author}
  {\bibfnamefont {J.}~\bibnamefont {Nunn}}, \ and\ \bibinfo {author}
  {\bibfnamefont {I.~A.}\ \bibnamefont {Walmsley}},\ }\href
  {http://link.aps.org/doi/10.1103/PhysRevB.91.205108} {\bibfield  {journal}
  {\bibinfo  {journal} {Physical Review B}\ }\textbf {\bibinfo {volume} {91}},\
  \bibinfo {pages} {205108} (\bibinfo {year} {2015})}\BibitemShut {NoStop}%
\bibitem [{\citenamefont {Suzuki}\ \emph {et~al.}(2016)\citenamefont {Suzuki},
  \citenamefont {Singh}, \citenamefont {Bayer}, \citenamefont {Ludwig},
  \citenamefont {Wieck},\ and\ \citenamefont {Cundiff}}]{Suzuki2016}%
  \BibitemOpen
  \bibfield  {author} {\bibinfo {author} {\bibfnamefont {T.}~\bibnamefont
  {Suzuki}}, \bibinfo {author} {\bibfnamefont {R.}~\bibnamefont {Singh}},
  \bibinfo {author} {\bibfnamefont {M.}~\bibnamefont {Bayer}}, \bibinfo
  {author} {\bibfnamefont {A.}~\bibnamefont {Ludwig}}, \bibinfo {author}
  {\bibfnamefont {A.~D.}\ \bibnamefont {Wieck}}, \ and\ \bibinfo {author}
  {\bibfnamefont {S.~T.}\ \bibnamefont {Cundiff}},\ }\href@noop {} {\bibfield
  {journal} {\bibinfo  {journal} {Physical Review Letters}\ }\textbf {\bibinfo
  {volume} {117}},\ \bibinfo {pages} {157402} (\bibinfo {year}
  {2016})}\BibitemShut {NoStop}%
\bibitem [{\citenamefont {Sun}(2005)}]{Liu2005}%
  \BibitemOpen
  \bibfield  {author} {\bibinfo {author} {\bibfnamefont {Y.~C.}\ \bibnamefont
  {Sun}},\ }in\ \href@noop {} {\emph {\bibinfo {booktitle} {Spectroscopic
  Properties of Rare Earths in Optical Materials}}},\ \bibinfo {editor} {edited
  by\ \bibinfo {editor} {\bibfnamefont {G.}~\bibnamefont {Liu}}\ and\ \bibinfo
  {editor} {\bibfnamefont {B.}~\bibnamefont {Jacquier}}}\ (\bibinfo
  {publisher} {Tsinghua University Press and Springer-Verlag},\ \bibinfo
  {address} {Berlin Heidelberg},\ \bibinfo {year} {2005})\ pp.\ \bibinfo
  {pages} {379 -- 429}\BibitemShut {NoStop}%
\bibitem [{\citenamefont {Duan}\ and\ \citenamefont {Kimble}(2004)}]{Duan2004}%
  \BibitemOpen
  \bibfield  {author} {\bibinfo {author} {\bibfnamefont {L.~M.}\ \bibnamefont
  {Duan}}\ and\ \bibinfo {author} {\bibfnamefont {H.~J.}\ \bibnamefont
  {Kimble}},\ }\href {\doibase 10.1103/PhysRevLett.92.127902} {\bibfield
  {journal} {\bibinfo  {journal} {Physical Review Letters}\ }\textbf {\bibinfo
  {volume} {92}},\ \bibinfo {pages} {127902} (\bibinfo {year}
  {2004})}\BibitemShut {NoStop}%
\bibitem [{\citenamefont {Dudin}\ and\ \citenamefont
  {Kuzmich}(2012)}]{Dudin2012}%
  \BibitemOpen
  \bibfield  {author} {\bibinfo {author} {\bibfnamefont {Y.~O.}\ \bibnamefont
  {Dudin}}\ and\ \bibinfo {author} {\bibfnamefont {A.}~\bibnamefont
  {Kuzmich}},\ }\href {\doibase 10.1126/science.1217901} {\bibfield  {journal}
  {\bibinfo  {journal} {Science}\ }\textbf {\bibinfo {volume} {336}},\ \bibinfo
  {pages} {887} (\bibinfo {year} {2012})}\BibitemShut {NoStop}%
\bibitem [{\citenamefont {Lukin}(2003)}]{Lukin2003}%
  \BibitemOpen
  \bibfield  {author} {\bibinfo {author} {\bibfnamefont {M.~D.}\ \bibnamefont
  {Lukin}},\ }\href {\doibase 10.1103/RevModPhys.75.457} {\bibfield  {journal}
  {\bibinfo  {journal} {Reviews of Modern Physics}\ }\textbf {\bibinfo {volume}
  {75}},\ \bibinfo {pages} {457} (\bibinfo {year} {2003})}\BibitemShut
  {NoStop}%
\bibitem [{\citenamefont {Reim}\ \emph {et~al.}(2010)\citenamefont {Reim},
  \citenamefont {Nunn}, \citenamefont {Lorenz}, \citenamefont {Sussman},
  \citenamefont {Lee}, \citenamefont {Langford}, \citenamefont {Jaksch},\ and\
  \citenamefont {Walmsley}}]{Reim2010}%
  \BibitemOpen
  \bibfield  {author} {\bibinfo {author} {\bibfnamefont {K.~F.}\ \bibnamefont
  {Reim}}, \bibinfo {author} {\bibfnamefont {J.}~\bibnamefont {Nunn}}, \bibinfo
  {author} {\bibfnamefont {V.~O.}\ \bibnamefont {Lorenz}}, \bibinfo {author}
  {\bibfnamefont {B.~J.}\ \bibnamefont {Sussman}}, \bibinfo {author}
  {\bibfnamefont {K.~C.}\ \bibnamefont {Lee}}, \bibinfo {author} {\bibfnamefont
  {N.~K.}\ \bibnamefont {Langford}}, \bibinfo {author} {\bibfnamefont
  {D.}~\bibnamefont {Jaksch}}, \ and\ \bibinfo {author} {\bibfnamefont {I.~A.}\
  \bibnamefont {Walmsley}},\ }\href@noop {} {\bibfield  {journal} {\bibinfo
  {journal} {Nature Photonics}\ }\textbf {\bibinfo {volume} {4}},\ \bibinfo
  {pages} {218} (\bibinfo {year} {2010})}\BibitemShut {NoStop}%
\bibitem [{\citenamefont {Zhou}\ \emph {et~al.}(2017)\citenamefont {Zhou},
  \citenamefont {Rasmita}, \citenamefont {Li}, \citenamefont {Xiong},
  \citenamefont {Aharonovich},\ and\ \citenamefont {Gao}}]{Zhou2017}%
  \BibitemOpen
  \bibfield  {author} {\bibinfo {author} {\bibfnamefont {Y.}~\bibnamefont
  {Zhou}}, \bibinfo {author} {\bibfnamefont {A.}~\bibnamefont {Rasmita}},
  \bibinfo {author} {\bibfnamefont {K.}~\bibnamefont {Li}}, \bibinfo {author}
  {\bibfnamefont {Q.}~\bibnamefont {Xiong}}, \bibinfo {author} {\bibfnamefont
  {I.}~\bibnamefont {Aharonovich}}, \ and\ \bibinfo {author} {\bibfnamefont
  {W.-B.}\ \bibnamefont {Gao}},\ }\href
  {http://www.nature.com/doifinder/10.1038/ncomms14451$\backslash$nhttp://arxiv.org/abs/1610.00882$\backslash$nhttp://dx.doi.org/10.1038/ncomms14451}
  {\bibfield  {journal} {\bibinfo  {journal} {Nature Communications}\ }\textbf
  {\bibinfo {volume} {8}},\ \bibinfo {pages} {14451} (\bibinfo {year}
  {2017})}\BibitemShut {NoStop}%
\bibitem [{\citenamefont {Dietrich}\ \emph {et~al.}(2014)\citenamefont
  {Dietrich}, \citenamefont {Jahnke}, \citenamefont {Binder}, \citenamefont
  {Teraji}, \citenamefont {Isoya}, \citenamefont {Rogers},\ and\ \citenamefont
  {Jelezko}}]{Dietrich2014}%
  \BibitemOpen
  \bibfield  {author} {\bibinfo {author} {\bibfnamefont {A.}~\bibnamefont
  {Dietrich}}, \bibinfo {author} {\bibfnamefont {K.~D.}\ \bibnamefont
  {Jahnke}}, \bibinfo {author} {\bibfnamefont {J.~M.}\ \bibnamefont {Binder}},
  \bibinfo {author} {\bibfnamefont {T.}~\bibnamefont {Teraji}}, \bibinfo
  {author} {\bibfnamefont {J.}~\bibnamefont {Isoya}}, \bibinfo {author}
  {\bibfnamefont {L.~J.}\ \bibnamefont {Rogers}}, \ and\ \bibinfo {author}
  {\bibfnamefont {F.}~\bibnamefont {Jelezko}},\ }\href@noop {} {\bibfield
  {journal} {\bibinfo  {journal} {New Journal of Physics}\ }\textbf {\bibinfo
  {volume} {16}},\ \bibinfo {pages} {113019} (\bibinfo {year}
  {2014})}\BibitemShut {NoStop}%
\bibitem [{sup()}]{supp}%
  \BibitemOpen
  \href@noop {} {\enquote {\bibinfo {title} {See supplemental material at [url
  will be inserted by publisher] for details on experimental setup, sample
  characterization and further experimental and theoretical results.}}\
  }\BibitemShut {NoStop}%
\bibitem [{\citenamefont {Vitanov}\ \emph {et~al.}(2017)\citenamefont
  {Vitanov}, \citenamefont {Rangelov}, \citenamefont {Shore},\ and\
  \citenamefont {Bergmann}}]{Vitanov2017}%
  \BibitemOpen
  \bibfield  {author} {\bibinfo {author} {\bibfnamefont {N.~V.}\ \bibnamefont
  {Vitanov}}, \bibinfo {author} {\bibfnamefont {A.~A.}\ \bibnamefont
  {Rangelov}}, \bibinfo {author} {\bibfnamefont {B.~W.}\ \bibnamefont {Shore}},
  \ and\ \bibinfo {author} {\bibfnamefont {K.}~\bibnamefont {Bergmann}},\
  }\href {\doibase 10.1103/RevModPhys.89.015006} {\bibfield  {journal}
  {\bibinfo  {journal} {Reviews of Modern Physics}\ }\textbf {\bibinfo {volume}
  {89}},\ \bibinfo {pages} {015006} (\bibinfo {year} {2017})}\BibitemShut
  {NoStop}%
\bibitem [{\citenamefont {Harris}\ \emph {et~al.}(1990)\citenamefont {Harris},
  \citenamefont {Field},\ and\ \citenamefont {Imamo\v{g}lu}}]{Harris1990}%
  \BibitemOpen
  \bibfield  {author} {\bibinfo {author} {\bibfnamefont {S.~E.}\ \bibnamefont
  {Harris}}, \bibinfo {author} {\bibfnamefont {J.~E.}\ \bibnamefont {Field}}, \
  and\ \bibinfo {author} {\bibfnamefont {A.}~\bibnamefont {Imamo\v{g}lu}},\
  }\href {\doibase 10.1103/PhysRevLett.64.1107} {\bibfield  {journal} {\bibinfo
   {journal} {Physical Review Letters}\ }\textbf {\bibinfo {volume} {64}},\
  \bibinfo {pages} {1107} (\bibinfo {year} {1990})}\BibitemShut {NoStop}%
\bibitem [{\citenamefont {Nunn}\ \emph {et~al.}(2017)\citenamefont {Nunn},
  \citenamefont {Munns}, \citenamefont {Thomas}, \citenamefont {Kaczmarek},
  \citenamefont {Qiu}, \citenamefont {Feizpour}, \citenamefont {Poem},
  \citenamefont {Brecht}, \citenamefont {Saunders}, \citenamefont {Ledingham},
  \citenamefont {Reddy}, \citenamefont {Raymer},\ and\ \citenamefont
  {Walmsley}}]{PhysRevA.96.012338}%
  \BibitemOpen
  \bibfield  {author} {\bibinfo {author} {\bibfnamefont {J.}~\bibnamefont
  {Nunn}}, \bibinfo {author} {\bibfnamefont {J.~H.~D.}\ \bibnamefont {Munns}},
  \bibinfo {author} {\bibfnamefont {S.}~\bibnamefont {Thomas}}, \bibinfo
  {author} {\bibfnamefont {K.~T.}\ \bibnamefont {Kaczmarek}}, \bibinfo {author}
  {\bibfnamefont {C.}~\bibnamefont {Qiu}}, \bibinfo {author} {\bibfnamefont
  {A.}~\bibnamefont {Feizpour}}, \bibinfo {author} {\bibfnamefont
  {E.}~\bibnamefont {Poem}}, \bibinfo {author} {\bibfnamefont {B.}~\bibnamefont
  {Brecht}}, \bibinfo {author} {\bibfnamefont {D.~J.}\ \bibnamefont
  {Saunders}}, \bibinfo {author} {\bibfnamefont {P.~M.}\ \bibnamefont
  {Ledingham}}, \bibinfo {author} {\bibfnamefont {D.~V.}\ \bibnamefont
  {Reddy}}, \bibinfo {author} {\bibfnamefont {M.~G.}\ \bibnamefont {Raymer}}, \
  and\ \bibinfo {author} {\bibfnamefont {I.~A.}\ \bibnamefont {Walmsley}},\
  }\href {\doibase 10.1103/PhysRevA.96.012338} {\bibfield  {journal} {\bibinfo
  {journal} {Phys. Rev. A}\ }\textbf {\bibinfo {volume} {96}},\ \bibinfo
  {pages} {012338} (\bibinfo {year} {2017})}\BibitemShut {NoStop}%
\bibitem [{\citenamefont {Belsley}\ \emph {et~al.}(1993)\citenamefont
  {Belsley}, \citenamefont {Smithey}, \citenamefont {Wedding},\ and\
  \citenamefont {Raymer}}]{Belsley1993}%
  \BibitemOpen
  \bibfield  {author} {\bibinfo {author} {\bibfnamefont {M.}~\bibnamefont
  {Belsley}}, \bibinfo {author} {\bibfnamefont {D.~T.}\ \bibnamefont
  {Smithey}}, \bibinfo {author} {\bibfnamefont {K.}~\bibnamefont {Wedding}}, \
  and\ \bibinfo {author} {\bibfnamefont {M.~G.}\ \bibnamefont {Raymer}},\
  }\href {\doibase 10.1103/PhysRevA.48.1514} {\bibfield  {journal} {\bibinfo
  {journal} {Physical Review A}\ }\textbf {\bibinfo {volume} {48}},\ \bibinfo
  {pages} {1514} (\bibinfo {year} {1993})}\BibitemShut {NoStop}%
\bibitem [{\citenamefont {Kosachiov}\ \emph {et~al.}(1992)\citenamefont
  {Kosachiov}, \citenamefont {Matisov},\ and\ \citenamefont
  {Rozhdestvensky}}]{Kosachiov1992}%
  \BibitemOpen
  \bibfield  {author} {\bibinfo {author} {\bibfnamefont {D.~V.}\ \bibnamefont
  {Kosachiov}}, \bibinfo {author} {\bibfnamefont {B.~G.}\ \bibnamefont
  {Matisov}}, \ and\ \bibinfo {author} {\bibfnamefont {Y.~V.}\ \bibnamefont
  {Rozhdestvensky}},\ }\href {\doibase 10.1088/0953-4075/25/11/005} {\bibfield
  {journal} {\bibinfo  {journal} {Journal of Physics B: Atomic, Molecular and
  Optical Physics}\ }\textbf {\bibinfo {volume} {25}},\ \bibinfo {pages} {2473}
  (\bibinfo {year} {1992})}\BibitemShut {NoStop}%
\bibitem [{\citenamefont {Lugani}\ \emph {et~al.}(2016)\citenamefont {Lugani},
  \citenamefont {Banerjee}, \citenamefont {Maynard}, \citenamefont {Neveu},
  \citenamefont {Xie}, \citenamefont {Ghosh}, \citenamefont {Bretenaker},\ and\
  \citenamefont {Goldfarb}}]{Lugani2016}%
  \BibitemOpen
  \bibfield  {author} {\bibinfo {author} {\bibfnamefont {J.}~\bibnamefont
  {Lugani}}, \bibinfo {author} {\bibfnamefont {C.}~\bibnamefont {Banerjee}},
  \bibinfo {author} {\bibfnamefont {M.-A.}\ \bibnamefont {Maynard}}, \bibinfo
  {author} {\bibfnamefont {P.}~\bibnamefont {Neveu}}, \bibinfo {author}
  {\bibfnamefont {W.}~\bibnamefont {Xie}}, \bibinfo {author} {\bibfnamefont
  {R.}~\bibnamefont {Ghosh}}, \bibinfo {author} {\bibfnamefont
  {F.}~\bibnamefont {Bretenaker}}, \ and\ \bibinfo {author} {\bibfnamefont
  {F.}~\bibnamefont {Goldfarb}},\ }\href {\doibase 10.1364/OL.41.004731}
  {\bibfield  {journal} {\bibinfo  {journal} {Opt. Lett.}\ }\textbf {\bibinfo
  {volume} {41}},\ \bibinfo {pages} {4731} (\bibinfo {year}
  {2016})}\BibitemShut {NoStop}%
\bibitem [{\citenamefont {Neveu}\ \emph {et~al.}()\citenamefont {Neveu},
  \citenamefont {Banerjee}, \citenamefont {Lugani}, \citenamefont {Bretenaker},
  \citenamefont {Brion},\ and\ \citenamefont {Goldfarb}}]{Neveu}%
  \BibitemOpen
  \bibfield  {author} {\bibinfo {author} {\bibfnamefont {P.}~\bibnamefont
  {Neveu}}, \bibinfo {author} {\bibfnamefont {C.}~\bibnamefont {Banerjee}},
  \bibinfo {author} {\bibfnamefont {J.}~\bibnamefont {Lugani}}, \bibinfo
  {author} {\bibfnamefont {F.}~\bibnamefont {Bretenaker}}, \bibinfo {author}
  {\bibfnamefont {E.}~\bibnamefont {Brion}}, \ and\ \bibinfo {author}
  {\bibfnamefont {F.}~\bibnamefont {Goldfarb}},\ }\href
  {https://arxiv.org/pdf/1803.05435.pdf} {\ }\Eprint
  {http://arxiv.org/abs/1803.05435} {arXiv:1803.05435} \BibitemShut {NoStop}%
\bibitem [{\citenamefont {Shahrokhshahi}\ \emph {et~al.}()\citenamefont
  {Shahrokhshahi}, \citenamefont {Sagona-Stophel}, \citenamefont {Jordaan},
  \citenamefont {Namazi},\ and\ \citenamefont {Figueroa}}]{Shahrokhshahi2018}%
  \BibitemOpen
  \bibfield  {author} {\bibinfo {author} {\bibfnamefont {R.}~\bibnamefont
  {Shahrokhshahi}}, \bibinfo {author} {\bibfnamefont {S.}~\bibnamefont
  {Sagona-Stophel}}, \bibinfo {author} {\bibfnamefont {B.}~\bibnamefont
  {Jordaan}}, \bibinfo {author} {\bibfnamefont {M.}~\bibnamefont {Namazi}}, \
  and\ \bibinfo {author} {\bibfnamefont {E.}~\bibnamefont {Figueroa}},\ }\href
  {http://arxiv.org/abs/1803.07012} {\ }\Eprint
  {http://arxiv.org/abs/1803.07012} {arXiv:1803.07012} \BibitemShut {NoStop}%
\bibitem [{\citenamefont {Gorshkov}\ \emph {et~al.}(2007)\citenamefont
  {Gorshkov}, \citenamefont {Andr\'e}, \citenamefont {Fleischhauer},
  \citenamefont {S\o{}rensen},\ and\ \citenamefont {Lukin}}]{gorshkov2007}%
  \BibitemOpen
  \bibfield  {author} {\bibinfo {author} {\bibfnamefont {A.~V.}\ \bibnamefont
  {Gorshkov}}, \bibinfo {author} {\bibfnamefont {A.}~\bibnamefont {Andr\'e}},
  \bibinfo {author} {\bibfnamefont {M.}~\bibnamefont {Fleischhauer}}, \bibinfo
  {author} {\bibfnamefont {A.~S.}\ \bibnamefont {S\o{}rensen}}, \ and\ \bibinfo
  {author} {\bibfnamefont {M.~D.}\ \bibnamefont {Lukin}},\ }\href {\doibase
  10.1103/PhysRevLett.98.123601} {\bibfield  {journal} {\bibinfo  {journal}
  {Phys. Rev. Lett.}\ }\textbf {\bibinfo {volume} {98}},\ \bibinfo {pages}
  {123601} (\bibinfo {year} {2007})}\BibitemShut {NoStop}%
\bibitem [{\citenamefont {Harris}\ and\ \citenamefont
  {Yamamoto}(1998)}]{harris1998}%
  \BibitemOpen
  \bibfield  {author} {\bibinfo {author} {\bibfnamefont {S.~E.}\ \bibnamefont
  {Harris}}\ and\ \bibinfo {author} {\bibfnamefont {Y.}~\bibnamefont
  {Yamamoto}},\ }\href {\doibase 10.1103/PhysRevLett.81.3611} {\bibfield
  {journal} {\bibinfo  {journal} {Phys. Rev. Lett.}\ }\textbf {\bibinfo
  {volume} {81}},\ \bibinfo {pages} {3611} (\bibinfo {year}
  {1998})}\BibitemShut {NoStop}%
\bibitem [{\citenamefont {Liu}\ \emph {et~al.}(2016)\citenamefont {Liu},
  \citenamefont {Chen}, \citenamefont {Chen}, \citenamefont {Lo}, \citenamefont
  {Tsai}, \citenamefont {Yu}, \citenamefont {Chen},\ and\ \citenamefont
  {Chen}}]{liu2016}%
  \BibitemOpen
  \bibfield  {author} {\bibinfo {author} {\bibfnamefont {Z.-Y.}\ \bibnamefont
  {Liu}}, \bibinfo {author} {\bibfnamefont {Y.-H.}\ \bibnamefont {Chen}},
  \bibinfo {author} {\bibfnamefont {Y.-C.}\ \bibnamefont {Chen}}, \bibinfo
  {author} {\bibfnamefont {H.-Y.}\ \bibnamefont {Lo}}, \bibinfo {author}
  {\bibfnamefont {P.-J.}\ \bibnamefont {Tsai}}, \bibinfo {author}
  {\bibfnamefont {I.~A.}\ \bibnamefont {Yu}}, \bibinfo {author} {\bibfnamefont
  {Y.-C.}\ \bibnamefont {Chen}}, \ and\ \bibinfo {author} {\bibfnamefont
  {Y.-F.}\ \bibnamefont {Chen}},\ }\href {\doibase
  10.1103/PhysRevLett.117.203601} {\bibfield  {journal} {\bibinfo  {journal}
  {Phys. Rev. Lett.}\ }\textbf {\bibinfo {volume} {117}},\ \bibinfo {pages}
  {203601} (\bibinfo {year} {2016})}\BibitemShut {NoStop}%
\bibitem [{\citenamefont {Tiarks}\ \emph {et~al.}(2016)\citenamefont {Tiarks},
  \citenamefont {Schmidt}, \citenamefont {Rempe},\ and\ \citenamefont
  {D{\"u}rr}}]{tiarks2016}%
  \BibitemOpen
  \bibfield  {author} {\bibinfo {author} {\bibfnamefont {D.}~\bibnamefont
  {Tiarks}}, \bibinfo {author} {\bibfnamefont {S.}~\bibnamefont {Schmidt}},
  \bibinfo {author} {\bibfnamefont {G.}~\bibnamefont {Rempe}}, \ and\ \bibinfo
  {author} {\bibfnamefont {S.}~\bibnamefont {D{\"u}rr}},\ }\href
  {http://advances.sciencemag.org/content/2/4/e1600036} {\bibfield  {journal}
  {\bibinfo  {journal} {Science Advances}\ }\textbf {\bibinfo {volume} {2}},\
  \bibinfo {pages} {e1600036} (\bibinfo {year} {2016})}\BibitemShut {NoStop}%
\bibitem [{\citenamefont {Beck}\ \emph {et~al.}(2016)\citenamefont {Beck},
  \citenamefont {Hosseini}, \citenamefont {Duan},\ and\ \citenamefont
  {Vuleti\'{c}}}]{beck2016}%
  \BibitemOpen
  \bibfield  {author} {\bibinfo {author} {\bibfnamefont {K.~M.}\ \bibnamefont
  {Beck}}, \bibinfo {author} {\bibfnamefont {M.}~\bibnamefont {Hosseini}},
  \bibinfo {author} {\bibfnamefont {Y.}~\bibnamefont {Duan}}, \ and\ \bibinfo
  {author} {\bibfnamefont {V.}~\bibnamefont {Vuleti\'{c}}},\ }\href {\doibase
  10.1073/pnas.1524117113} {\bibfield  {journal} {\bibinfo  {journal}
  {Proceedings of the National Academy of Sciences}\ }\textbf {\bibinfo
  {volume} {113}},\ \bibinfo {pages} {9740} (\bibinfo {year}
  {2016})}\BibitemShut {NoStop}%
\bibitem [{\citenamefont {Iwasaki}\ \emph {et~al.}(2015)\citenamefont
  {Iwasaki}, \citenamefont {Ishibashi}, \citenamefont {Miyamoto}, \citenamefont
  {Doi}, \citenamefont {Kobayashi}, \citenamefont {Miyazaki}, \citenamefont
  {Tahara}, \citenamefont {Jahnke}, \citenamefont {Rogers}, \citenamefont
  {Naydenov}, \citenamefont {Jelezko}, \citenamefont {Yamasaki}, \citenamefont
  {Nagamachi}, \citenamefont {Inubushi}, \citenamefont {Mizuochi},\ and\
  \citenamefont {Hatano}}]{Iwasaki2015}%
  \BibitemOpen
  \bibfield  {author} {\bibinfo {author} {\bibfnamefont {T.}~\bibnamefont
  {Iwasaki}}, \bibinfo {author} {\bibfnamefont {F.}~\bibnamefont {Ishibashi}},
  \bibinfo {author} {\bibfnamefont {Y.}~\bibnamefont {Miyamoto}}, \bibinfo
  {author} {\bibfnamefont {Y.}~\bibnamefont {Doi}}, \bibinfo {author}
  {\bibfnamefont {S.}~\bibnamefont {Kobayashi}}, \bibinfo {author}
  {\bibfnamefont {T.}~\bibnamefont {Miyazaki}}, \bibinfo {author}
  {\bibfnamefont {K.}~\bibnamefont {Tahara}}, \bibinfo {author} {\bibfnamefont
  {K.~D.}\ \bibnamefont {Jahnke}}, \bibinfo {author} {\bibfnamefont {L.~J.}\
  \bibnamefont {Rogers}}, \bibinfo {author} {\bibfnamefont {B.}~\bibnamefont
  {Naydenov}}, \bibinfo {author} {\bibfnamefont {F.}~\bibnamefont {Jelezko}},
  \bibinfo {author} {\bibfnamefont {S.}~\bibnamefont {Yamasaki}}, \bibinfo
  {author} {\bibfnamefont {S.}~\bibnamefont {Nagamachi}}, \bibinfo {author}
  {\bibfnamefont {T.}~\bibnamefont {Inubushi}}, \bibinfo {author}
  {\bibfnamefont {N.}~\bibnamefont {Mizuochi}}, \ and\ \bibinfo {author}
  {\bibfnamefont {M.}~\bibnamefont {Hatano}},\ }\href {\doibase
  10.1038/srep12882} {\bibfield  {journal} {\bibinfo  {journal} {Scientific
  Reports}\ }\textbf {\bibinfo {volume} {5}},\ \bibinfo {pages} {12882}
  (\bibinfo {year} {2015})}\BibitemShut {NoStop}%
\bibitem [{\citenamefont {Iwasaki}\ \emph {et~al.}(2017)\citenamefont
  {Iwasaki}, \citenamefont {Miyamoto}, \citenamefont {Taniguchi}, \citenamefont
  {Siyushev}, \citenamefont {Metsch}, \citenamefont {Jelezko},\ and\
  \citenamefont {Hatano}}]{Iwasaki2017}%
  \BibitemOpen
  \bibfield  {author} {\bibinfo {author} {\bibfnamefont {T.}~\bibnamefont
  {Iwasaki}}, \bibinfo {author} {\bibfnamefont {Y.}~\bibnamefont {Miyamoto}},
  \bibinfo {author} {\bibfnamefont {T.}~\bibnamefont {Taniguchi}}, \bibinfo
  {author} {\bibfnamefont {P.}~\bibnamefont {Siyushev}}, \bibinfo {author}
  {\bibfnamefont {M.~H.}\ \bibnamefont {Metsch}}, \bibinfo {author}
  {\bibfnamefont {F.}~\bibnamefont {Jelezko}}, \ and\ \bibinfo {author}
  {\bibfnamefont {M.}~\bibnamefont {Hatano}},\ }\href {\doibase
  10.1103/PhysRevLett.119.253601} {\bibfield  {journal} {\bibinfo  {journal}
  {Physical Review Letters}\ }\textbf {\bibinfo {volume} {119}},\ \bibinfo
  {pages} {253601} (\bibinfo {year} {2017})}\BibitemShut {NoStop}%
\bibitem [{\citenamefont {Larico}\ \emph {et~al.}(2009)\citenamefont {Larico},
  \citenamefont {Justo}, \citenamefont {Machado},\ and\ \citenamefont
  {Assali}}]{Larico2009}%
  \BibitemOpen
  \bibfield  {author} {\bibinfo {author} {\bibfnamefont {R.}~\bibnamefont
  {Larico}}, \bibinfo {author} {\bibfnamefont {J.~F.}\ \bibnamefont {Justo}},
  \bibinfo {author} {\bibfnamefont {W.~V.~M.}\ \bibnamefont {Machado}}, \ and\
  \bibinfo {author} {\bibfnamefont {L.~V.~C.}\ \bibnamefont {Assali}},\ }\href
  {\doibase 10.1103/PhysRevB.79.115202} {\bibfield  {journal} {\bibinfo
  {journal} {Physical Review B - Condensed Matter and Materials Physics}\
  }\textbf {\bibinfo {volume} {79}},\ \bibinfo {pages} {115202} (\bibinfo
  {year} {2009})}\BibitemShut {NoStop}%
\end{thebibliography}
\end{document}